\documentclass[doublecol]{epl2} 

\usepackage{amsmath}
\usepackage{mathtools}
\usepackage{amsfonts}
\usepackage{amssymb}
\usepackage{graphicx}
\usepackage{lmodern}
\usepackage{cite}
\usepackage{hyperref}
\usepackage{dcolumn}
\usepackage{bm}
\usepackage[mathlines]{lineno}
\usepackage{float}
\usepackage[T1]{fontenc}
\usepackage{subfigure}

\title{Deviation from $\mu-\tau$ Symmetry with New Approaches}

\author{Pralay Chakraborty\inst{1} \and Tanmay Dev\inst{1} \and Subhankar Roy\inst{1}}

\institute{                    
  \inst{1} Department of Physics, Gauhati University, India, 781014
}

\abstract{We propose two new predictive neutrino mass matrix textures that deviate from $\mu$-$\tau$ symmetry and explore their phenomenological implications. These textures are significant due to their predictive nature and the unique approach they introduce for breaking the $\mu$-$\tau$ symmetry. Both textures predict six physical parameters and impose sharp constraints on $\theta_{23}$ and $\delta$, considering both normal and inverted neutrino mass orderings. In addition, we realize the textures in their exact form within the framework of the seesaw mechanism, based on the $SU(2)_L \times A_4 \times Z_3$ symmetry group.}

\begin{document}

\maketitle

\newcolumntype{P}[1]{>{\centering\arraybackslash}p{#1}}

\section{Introduction}

In the year 1957, Bruno Pontecorvo proposed the idea of neutrino oscillation, suggesting that neutrinos change their identity as they move through space \cite{Pontecorvo:1957cp}. This phenomenon arises from the fact that the three flavours of neutrinos ($\nu_{l=e,\mu,\tau}$) are  admixtures of three mass eigenstates $(\nu_{i=1,2,3})$ with corresponding mass values ($m_{i=1,2,3}$). The phenomenon of neutrino oscillation predicts that neutrinos must have non-zero mass, which contradicts with the predictions of the Standard Model\,(SM)\,\cite{Glashow:1961tr, Weinberg:1967tq, Salam:1968rm}. To understand the origin of neutrino mass, it is necessary to explore theories that go beyond the SM. The neutrino mass matrix ($M_\nu$) originates from the Yukawa Lagrangian, formulated within the framework of the see-saw mechanism \cite{ Akhmedov:1999tm, Schechter:1980gr}. The $M_\nu$ contains twelve parameters: three mass eigenvalues ($m_{i=1,2,3}$), three mixing angles ($\theta_{12}, \theta_{13}, \theta_{23}$), three CP phases ($\delta, \alpha, \beta$), and three unphysical phases ($\phi_1, \phi_2, \phi_3$). Recent experiments estimate three mixing angles and two mass-squared differences\,($\Delta\,m_{21}^2$ and $\Delta\,m_{31}^2$) precisely \cite{Esteban:2024eli}. However, the neutrino mass ordering, the octant of $\theta_{23}$ and a precise bound for $\delta$ are not determined yet. In addition, the three individual mass eigenvalues and the Majorana phases are not witnessed by the oscillation experiments. 

However, the Majorana neutrino mass matrix shelters the information of all nine physical parameters. In this regard, by imposing certain constraints on $M_\nu$, some of the physical parameters can be estimated. In the literature, various constraints, such as texture zeroes \cite{Ludl:2014axa}, vanishing minors\,\cite{Lashin:2009yd}, hybrid textures \cite{Liu:2013oxa} and $\mu-\tau$ symmetry \cite{Harrison:2002er} have been applied to $M_\nu$ so that some physical parameters can be predicted. In addition, we discuss some new constraints in Ref.\,\cite{Chakraborty:2022ess, Chakraborty:2024hhq}. It is worth noting that experiments have ruled out the $\mu-\tau$ symmetry due to its prediction of $\theta_{13}$ being zero \cite{Esteban:2024eli}. Despite this, the $\mu-\tau$ symmetry along with its predictions has been well-established, the theorists have made several attempts to explore deviations from the exact $\mu-\tau$ symmetry in the form of $\mu-\tau$ anti-symmetry \cite{Xing:2015fdg}, $\mu-\tau$ reflection symmetry \cite{Xing:2022uax, Vien:2024cnm, Vien:2024seg}, $\mu-\tau$ mixed symmetry \cite{Dey:2022qpu, Chakraborty:2023msb}. In the present work, we deviate from the existing literature and adopt a new approach to break exact $\mu-\tau$ symmetry. We propose two neutrino mass matrix textures as shown in the following,

\begin{eqnarray}
\label{M1}
M^1_\nu &=&
\begin{bmatrix}
A &  B  &   (i B)^* \\
B &  2B^*  &  D \\
(i B)^* & D &  -2B \\
\end{bmatrix},\\
\label{M2} M^2_\nu &=&
\begin{bmatrix}
A &  -B  &  -i B/2 \\
-B &  B  &  D \\
-i B/2 & D &  2B \\
\end{bmatrix},
\end{eqnarray}

where, $A,\,B$ and $D$ appearing in both the textures are complex parameters. Both textures introduce new constraints on the neutrino mass matrix that have not been studied earlier. Unlike traditional approaches, the proposed textures are more phenomenologically motivated, as they predict six physical parameters. It is to be noted that the matrices as appearing in Eqs. (\ref{M1}) may appear arbitrary and realizing these textures from symmetry is quite challenging. However, it is essential to explore their origins from symmetry frameworks to establish a theoretical foundation from first principles. In this regard, we realize the proposed textures from $SU(2)_L \times A_4 \times Z_3$ symmetry group in the light of seesaw mechanism.

\section{Formalism}

In addition to $M_{\nu}$, the neutrino physics phenomenology deals with an important quantity known as the Pontecorvo-Maki-Nakagawa-Sakata (PMNS) matrix \cite{Maki:1962mu}, $V$, which is a $3 \times 3$ unitary matrix. In general, the PMNS matrix is defined in a basis where the charged lepton mass matrix is diagonal (flavour basis). The Particle Data Group \cite{ParticleDataGroup:2020ssz} has adopted a parametrization scheme for $V$ as shown in the following,

\begin{equation}
\label{pmns}
V= P_{\phi}. U. P_M, \nonumber
\end{equation}

where, $P_{\phi} =diag (e^{i\phi_1},e^{i\phi_2},e^{i\phi_3})$ and $P_M=diag(e^{i\alpha},e^{i\beta},1)$. The matrix $U$ is depicted as shown below,

\begin{eqnarray}
U &=& \begin{bmatrix}
1 & 0 & 0\\
0 & c_{23} & s_{23}\\
0 & - s_{23} & c_{23}
\end{bmatrix}\times \begin{bmatrix}
c_{13} & 0 & s_{13}\,e^{-i\delta}\\
0 & 1 & 0\\
-s_{13} e^{i\delta} & 0 & c_{13}
\end{bmatrix}\nonumber\\
&& \quad\quad\times\begin{bmatrix}
c_{12} & s_{12} & 0\\
-s_{12} & c_{12} & 0\\
0 & 0 & 1
\end{bmatrix},
\end{eqnarray}

where, $s_{ij}=\sin\theta_{ij}$ and $c_{ij}=\cos\theta_{ij}$. 

\begin{figure*}
  \centering
    \subfigure[]{\includegraphics[width=0.24\textwidth]{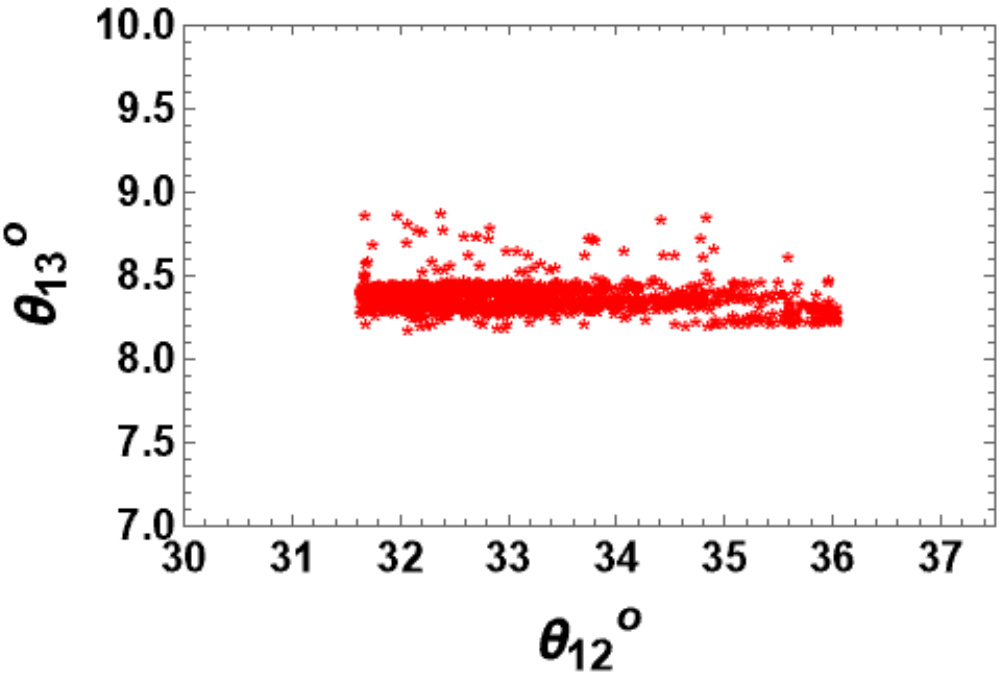}\label{fig:1(a)}} 
    \subfigure[]{\includegraphics[width=0.24\textwidth]{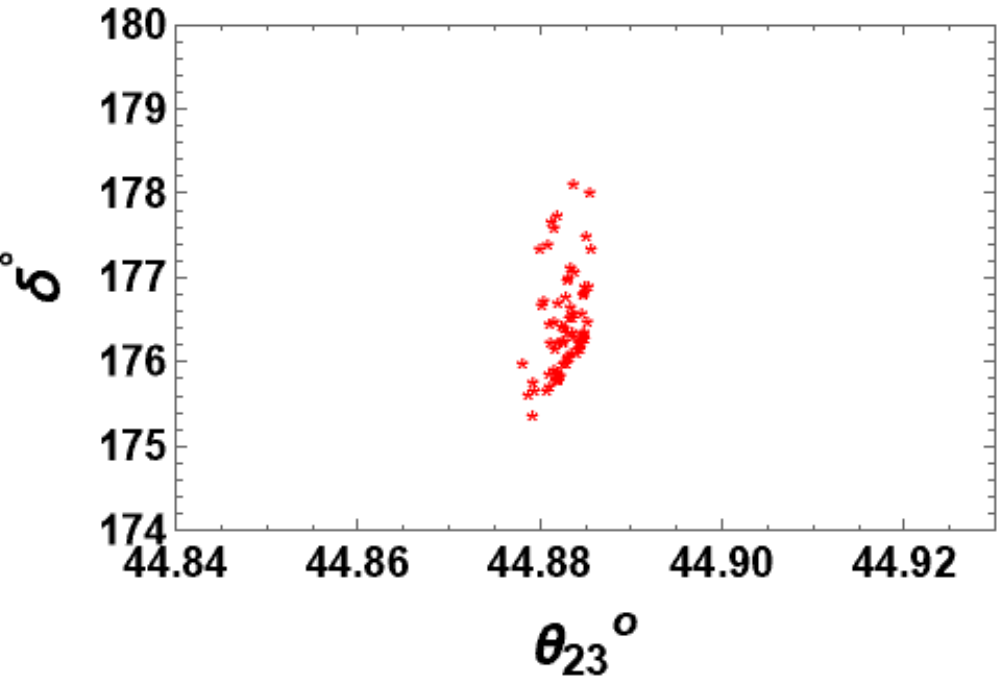}\label{fig:1(b)}} 
    \subfigure[]{\includegraphics[width=0.24\textwidth]{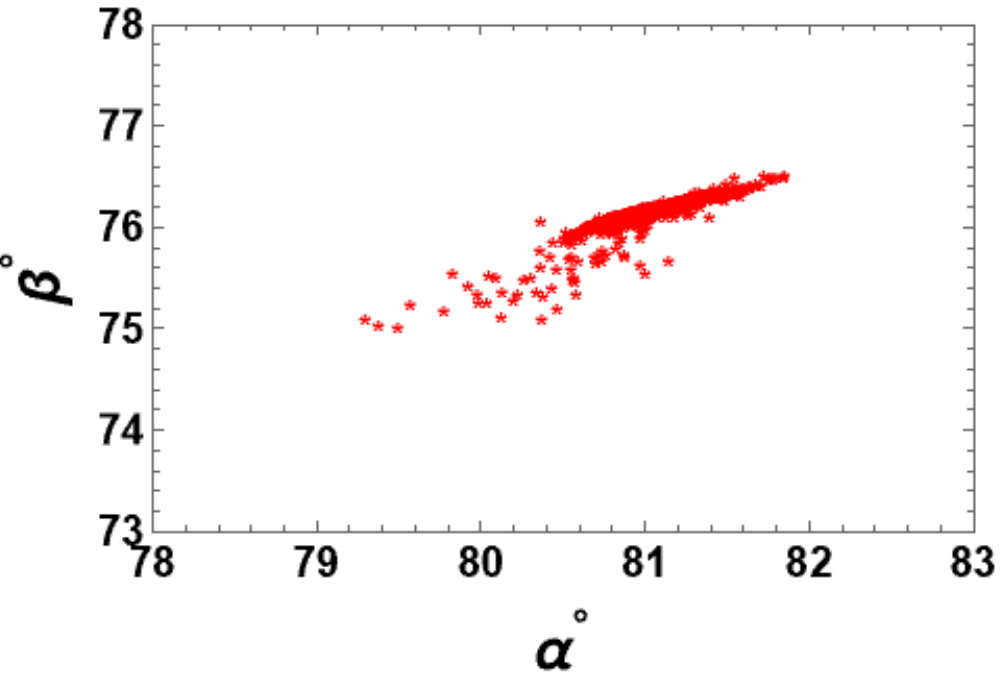}\label{fig:1(c)}}
    \subfigure[]{\includegraphics[width=0.24\textwidth]{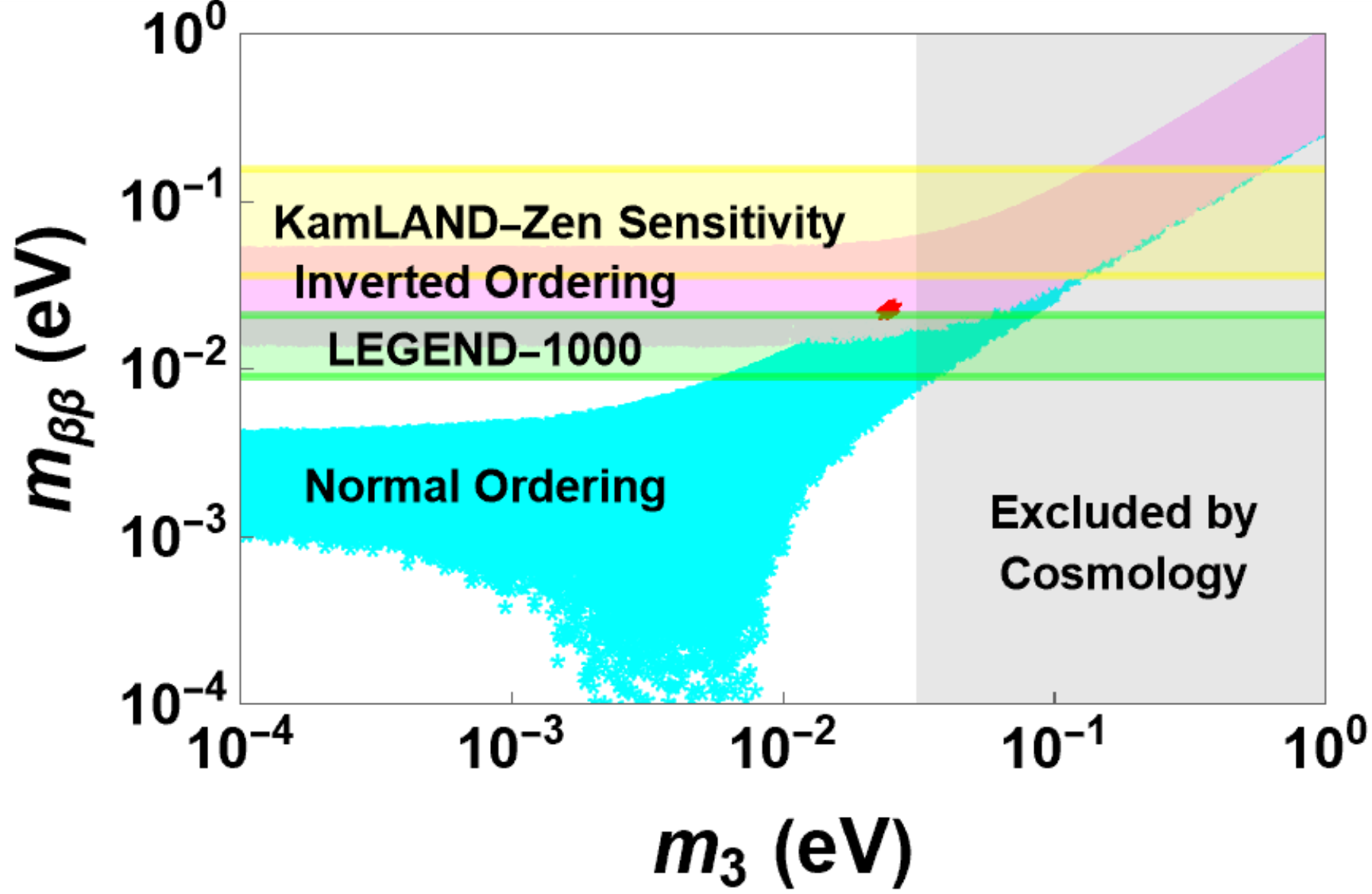}\label{fig:1(d)}} 
    \caption{The correlation plots between (a) $\theta_{12}$ and $\theta_{13}$, (b) $\theta_{23}$ and $\delta$, (c)  $\alpha$ and $\beta$ for the texture $M_{\nu}^1$ under normal ordering of neutrino masses, (d) Shows the variation of $m_{\beta\beta}$ against the lightest neutrino mass ($m_1$).}
\label{fig:1}
\end{figure*}

As discussed in Section 1, $M_\nu$ shelters information about all physical parameters. To extract this information, we need to diagonalize $M_\nu$: $M_\nu^d=V^T M_\nu V$. It is to be noted that the unphysical phases ($\phi_{i=1,2,3}$) can be removed from $V$ by redefining the charged lepton fields in terms of these phases. However, in the present work, we stick to the general form of $M_\nu$ containing the unphysical phases. We distinguish $M_\nu$ from $\tilde{M_\nu}$ that excludes the unphysical phases as shown in the following,

\begin{equation}
M_\nu = P^*_\phi \tilde{M_\nu} P^*_\phi, 
\label{unphysical phases transformation}
\end{equation}

where $P_\phi$ carries the unphysical phases.

The constrained relations appearing in $M_{\nu}^1$ as shown in Eq.\,(\ref{M1}) hold good provided the following conditions for the unphysical phases are satisfied,

\begin{eqnarray}
2\phi_1+\phi_2+\phi_3 &=& Arg[\tilde{(M^2_\nu)}_{12}]+Arg[\tilde{(M^2_\nu)}_{13}]+\frac{\pi}{2}, \nonumber\\
\phi_2+\phi_3 &=& \frac{1}{2}\left( Arg[\tilde{(M^2_\nu)}_{22}]+Arg[\tilde{(M^2_\nu)}_{33}-\frac{\pi}{2}\right), \nonumber\\
\phi_1+\phi_2-2\phi_3 &=&  Arg[\tilde{(M^2_\nu)}_{12}]-Arg[\tilde{(M^2_\nu)}_{33}-\frac{\pi}{2}. \nonumber
\end{eqnarray}

On the other hand, for $M_{\nu}^2$ as appearing in Eq.\,(\ref{M2}), the conditions for the unphysical phases are given by,

\begin{eqnarray}
\phi_2-\phi_3 &=& Arg[\tilde{(M^1_\nu)}_{12}]-Arg[\tilde{(M^1_\nu)}_{13}]+\frac{\pi}{2}, \nonumber\\
\phi_2-\phi_3 &=& \frac{1}{2}\left( Arg[\tilde{(M^1_\nu)}_{22}]-Arg[\tilde{(M^1_\nu)}_{33}\right), \nonumber\\
\phi_1-\phi_2 &=&  Arg[\tilde{(M^1_\nu)}_{12}]-Arg[\tilde{(M^1_\nu)}_{22}-\frac{\pi}{2}. \nonumber
\end{eqnarray}

It is to be noted that the choice between sticking to $M_\nu$ or $\tilde{M_\nu}$ is subjective. However, keeping the said phases intact in $M_\nu$ provides some phenomenological prospects. The importance of these unphysical phases is highlighted in the Ref.\,\cite{Xing:2022uax}. It is evident from Eq.\,(\ref{unphysical phases transformation}) that $\tilde{M_{\nu}}$ does not involve the unphysical phases. In this regard, constraints applied to $\tilde{M_{\nu}}$ ensure that predictions of physical parameters remain unaffected by these phases. However, the same is not true if we are dealing with $M_{\nu}$. In most studies, model builders prefer to stick to $\tilde{M_{\nu}}$, along with the fact that $\phi_1$, $\phi_2$, and $\phi_3$ are removed by redefining the charged lepton fields. However, there are no such theoretical impositions that actually dictate the removal of the said phases. In the present work, we try to formulate constraints on the mass matrix elements of $M_{\nu}$ rather than $\tilde{M_{\nu}}$.

\section{Numerical analysis}

  In this section, we discuss the phenomenological implications of the proposed textures $M^1_\nu$ and $M^2_\nu$ in the light of experimental observations. Both the textures shelters three complex correlations which in turn leads to six real transcendental equations: $f_i(\theta_{12}, \theta_{13}, \theta_{23}, m_1, m_2, m_3, \delta, \alpha, \beta, \phi_1, \phi_2, \phi_3)=0$, where, $i=1,2.3,4,5,6$.

  \begin{table}
\centering
\begin{tabular}{ ccccc}
\hline
\hline
 & NO  & NO  & IO  & IO  \\
Parameters & (Min) &  (Max) & (MIN) &  (Max)\\
\hline
\hline
$\theta_{12} /^\circ$ & 31.643 & 35.941 & 31.65 & 35.94\\
\hline
$\theta_{13} /^\circ$ & 8.186 & 8.87 & 8.45 & 8.87\\
\hline
$\theta_{23} /^\circ$ & 44.86 & 44.89 & 44.87 & 44.91\\
\hline
$\delta /^\circ$ & 173.53 & 178.44 & 338.22 & 339.48\\
\hline
$\alpha/^\circ$ & 79.29 & 81.85  & 66.45 & 67.61\\
\hline
$\beta/^\circ$ & 75.03 & 76.53  & 47.19 & 48.28\\
\hline
\end{tabular}
\caption{Shows the maximum and minimum values of $\theta_{12}$, $\theta_{13}$, $\theta_{23}$, $\delta$, $\alpha$ and $\beta$ for texture $M_{\nu}^1$ in the light of both normal and inverted orderings.}
\label{values of physical parameters 1}
\end{table} 

\begin{table}
\centering
\begin{tabular}{p{2cm} p{2cm} p{2cm}}
\hline
\hline
Parameters & Minimum Value &  Maximum Value \\
\hline
\hline
$\theta_{12} /^\circ$ & 31.64 & 35.92 \\
\hline
$\theta_{13} /^\circ$ & 8.32 & 8.85\\
\hline
$\theta_{23} /^\circ$ & 41.60 & 42.23 \\
\hline
$\delta /^\circ$ & 215.28 & 223.24 \\
\hline
$\alpha/^\circ$ & 107.70 & 111.81  \\
\hline
$\beta/^\circ$ & 265.59 & 267.87  \\
\hline
\end{tabular}
\caption{Shows the maximum and minimum values of $\theta_{12}$, $\theta_{13}$, $\theta_{23}$, $\delta$, $\alpha$ and $\beta$ for texture $M_{\nu}^2$ in the light of normal ordering.}
\label{values of physical parameters 2}
\end{table}
  
  The proposed texture $M^1_\nu$ deals with three constrained relations as shown in the following,
  
  \begin{eqnarray}
(M^1_{\nu})_{12} &=& (i (M^1_{\nu})_{13})^*, \label{first relation second texture}\\
(M^1_{\nu})_{22} &=& -((M^1_{\nu})_{33})^* ,\label{second relation second texture}\\
(M^1_{\nu})_{33} &=& -2((M^1_{\nu})_{12}) .\label{third relation second texture}
\end{eqnarray}

\begin{figure*}
  \centering
    \subfigure[]{\includegraphics[width=0.24\textwidth]{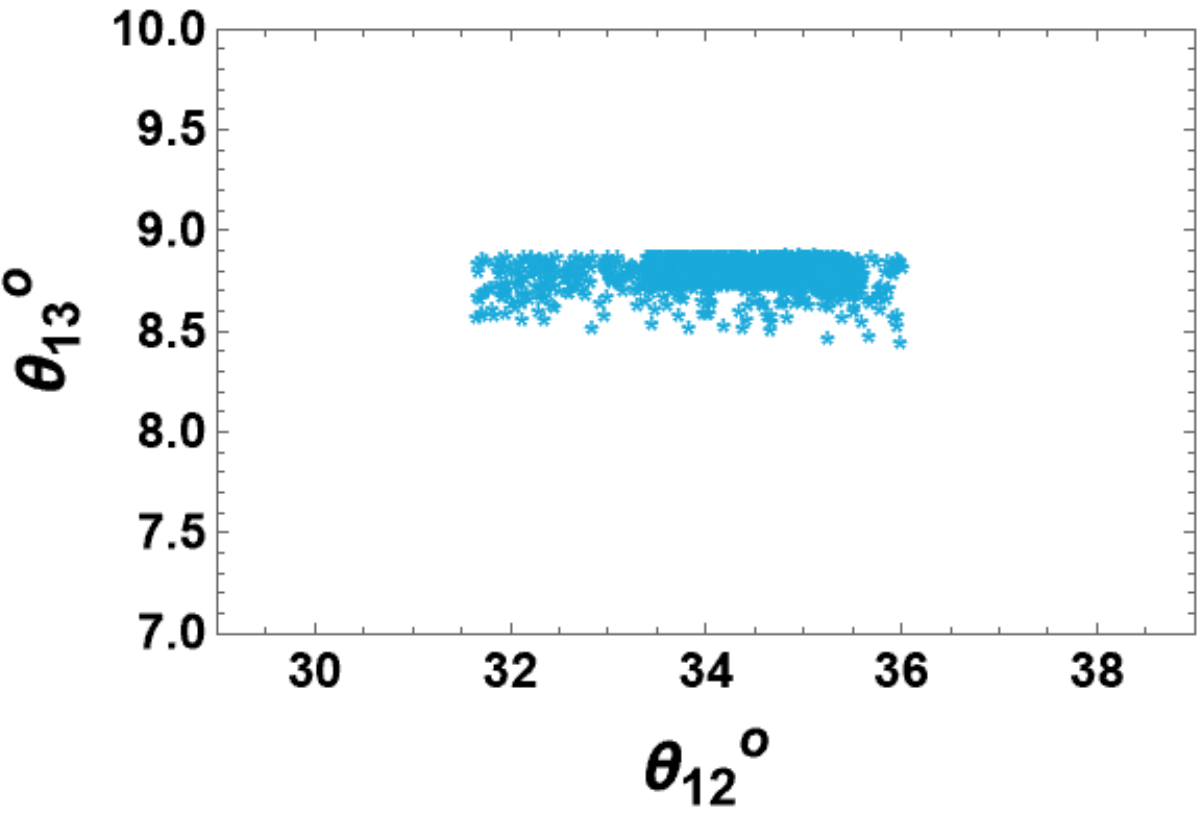}\label{fig:2(a)}} 
    \subfigure[]{\includegraphics[width=0.24\textwidth]{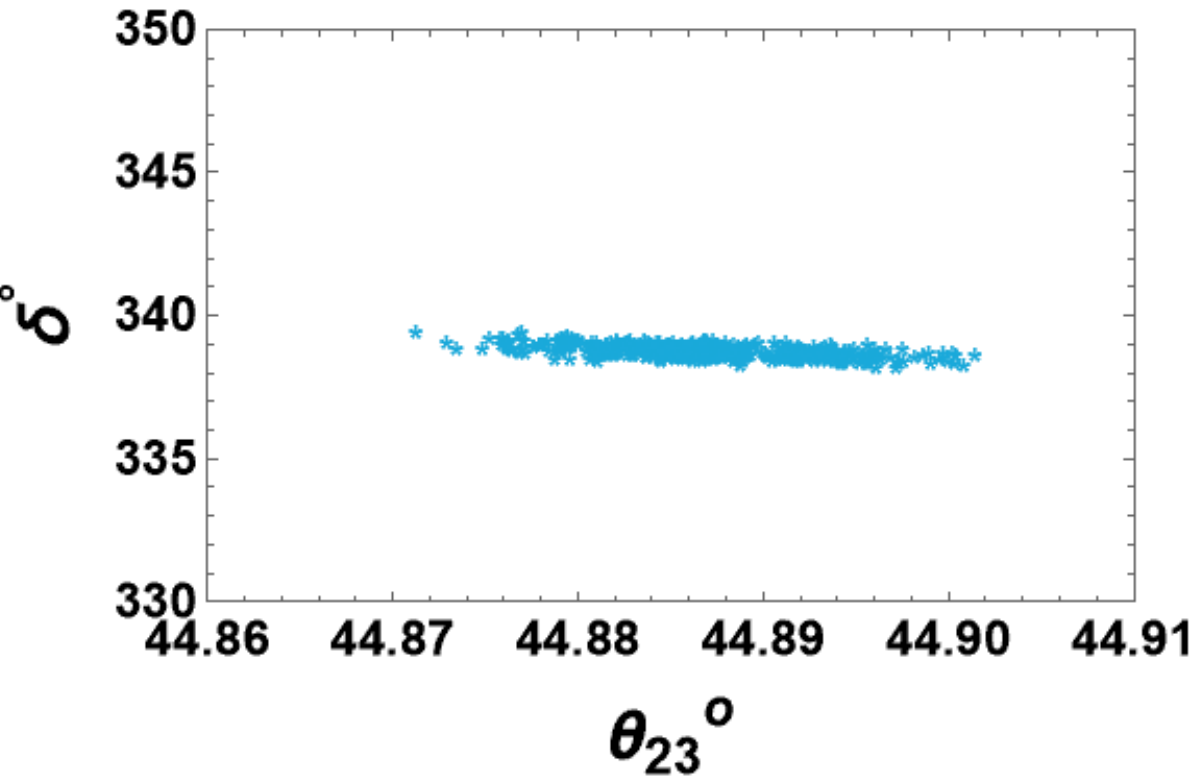}\label{fig:2(b)}} 
    \subfigure[]{\includegraphics[width=0.24\textwidth]{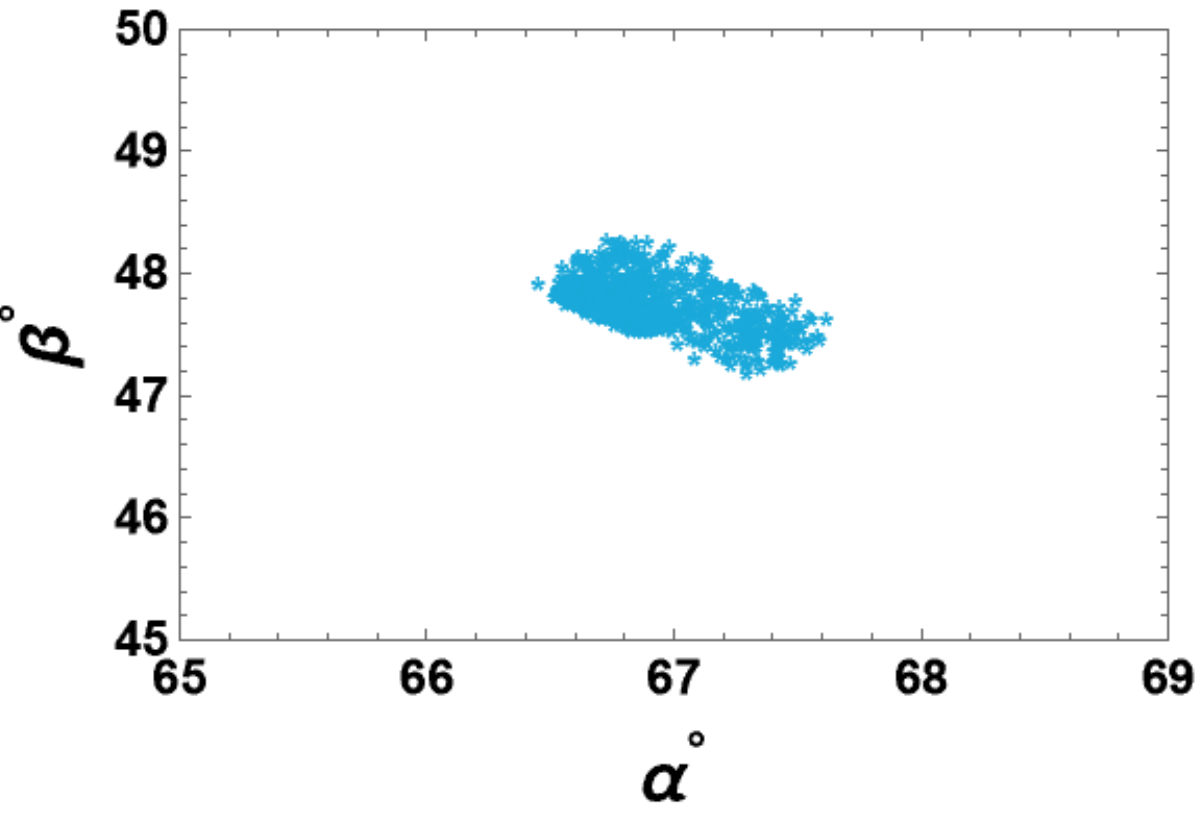}\label{fig:2(c)}}
    \subfigure[]{\includegraphics[width=0.25\textwidth]{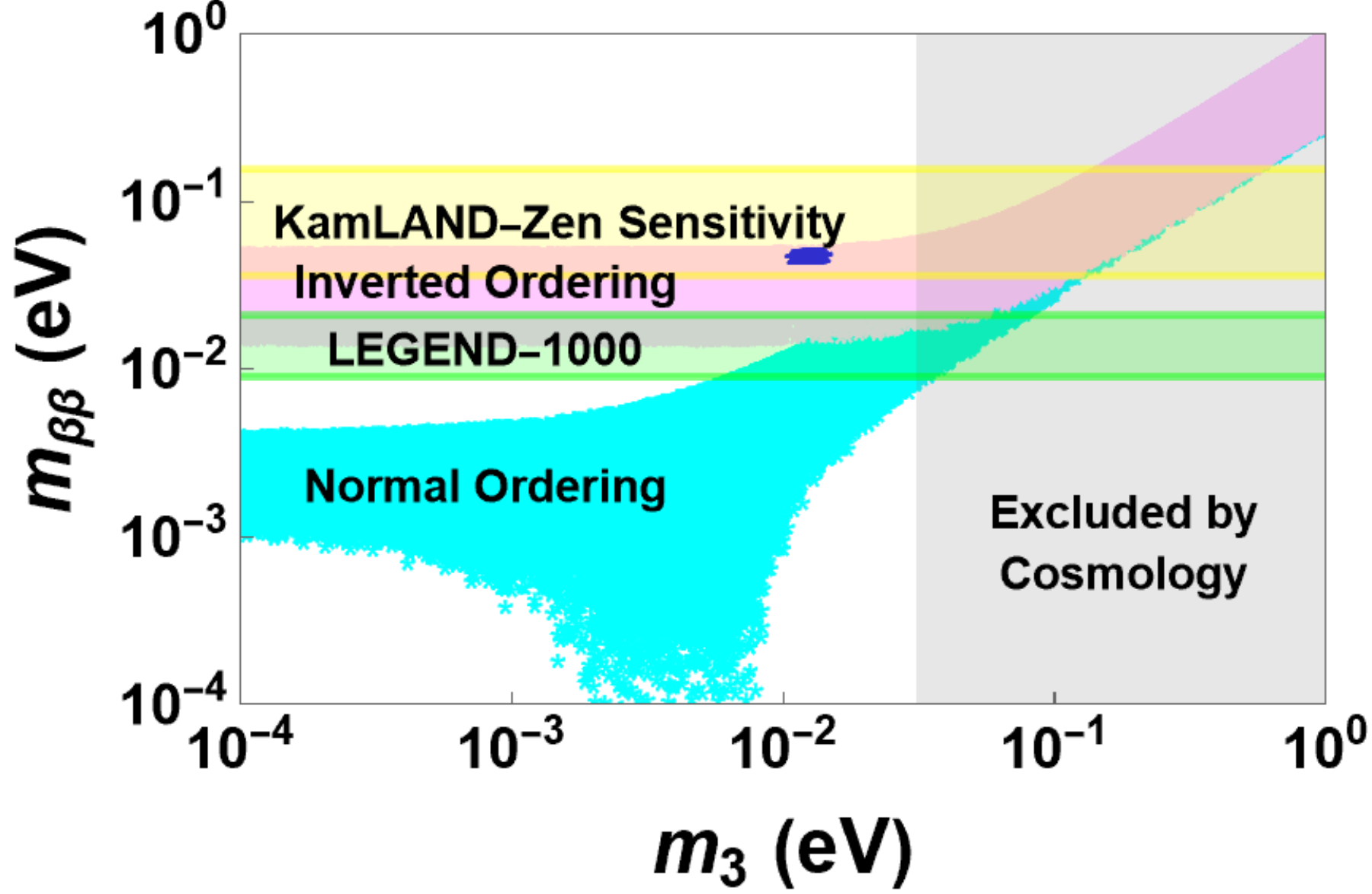}\label{fig:2(d)}} 
    \caption{The correlation plots between (a) $\theta_{12}$ and $\theta_{13}$, (b) $\theta_{23}$ and $\delta$, (c)  $\alpha$ and $\beta$ for the texture $M_{\nu}^1$ under inverted ordering of neutrino masses, (d) Shows the variation of $m_{\beta\beta}$ against the lightest neutrino mass ($m_3$).}
\label{fig:2}
\end{figure*}

The fourth relation $(M^1_{\nu})_{22}=2(M^1_{\nu})_{12}^*$ is not independent and can be derived from the above three. We exploit the said equations numerically and generate the sufficient number of random numbers for the parameters $\theta_{12}$, $\theta_{13}$, $\theta_{23}$, $\delta$, $\alpha$ and $\beta$ and obtain the correlations plots. We observe that the mixing scheme obtained from $M_{\nu}^1$ is consistent with the experiments under both normal and inverted orderings\,(see Figs.\,\ref{fig:1}-\ref{fig:2}). It is seen that both the textures constrain $\theta_{23}$ in the lower octant and $\delta$ in a sharp bound. In addition, the Majorana phases are constrained below $90^\circ$. The minimum and maximum values of the studied parameters in the light of both normal and inverted orderings are given in Table\,\ref{values of physical parameters 1}.

The texture $M^2_\nu$ shelters three constrained relations as shown below,

\begin{eqnarray}
(M^2_{\nu})_{12} &=& 2 i (M^2_{\nu})_{13}, \label{first relation first texture}\\
(M^2_{\nu})_{22} &=&\frac{(M^2_{\nu})_{33}}{2},  \label{second relation first texture}\\
(M^2_{\nu})_{22} &=&-(M^2_{\nu})_{12}.  \label{third relation first texture}
\end{eqnarray}

\begin{figure*}
  \centering
    \subfigure[]{\includegraphics[width=0.24\textwidth]{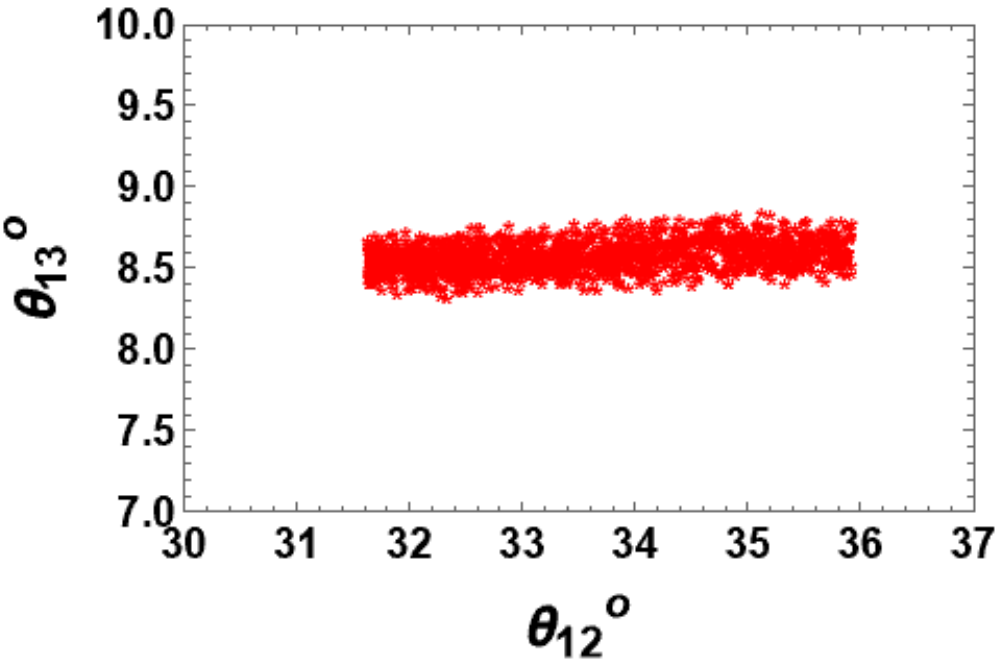}\label{fig:3(a)}} 
    \subfigure[]{\includegraphics[width=0.24\textwidth]{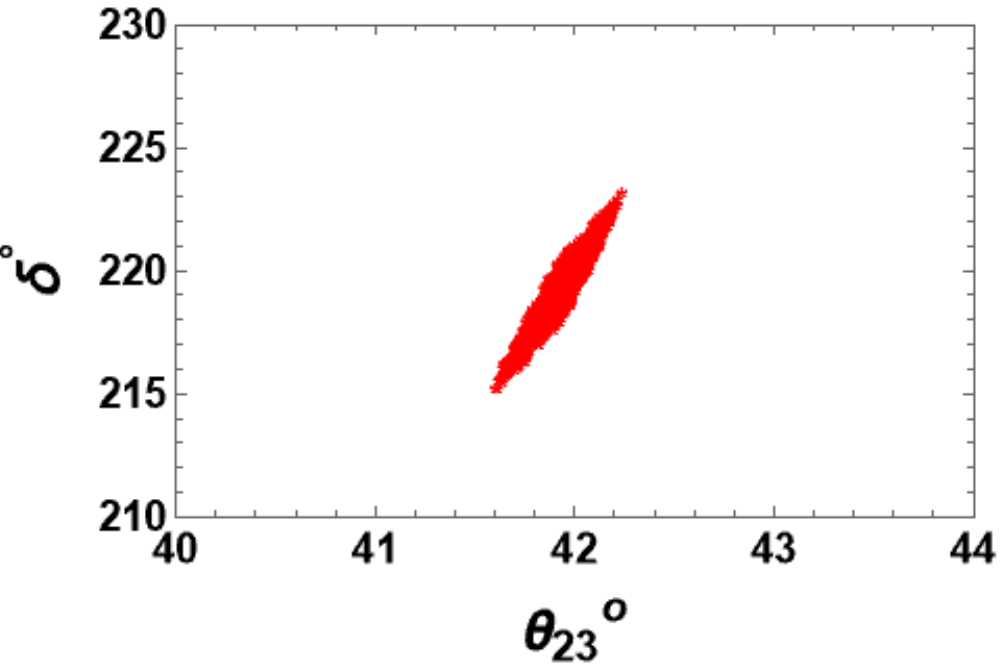}\label{fig:3(b)}} 
    \subfigure[]{\includegraphics[width=0.24\textwidth]{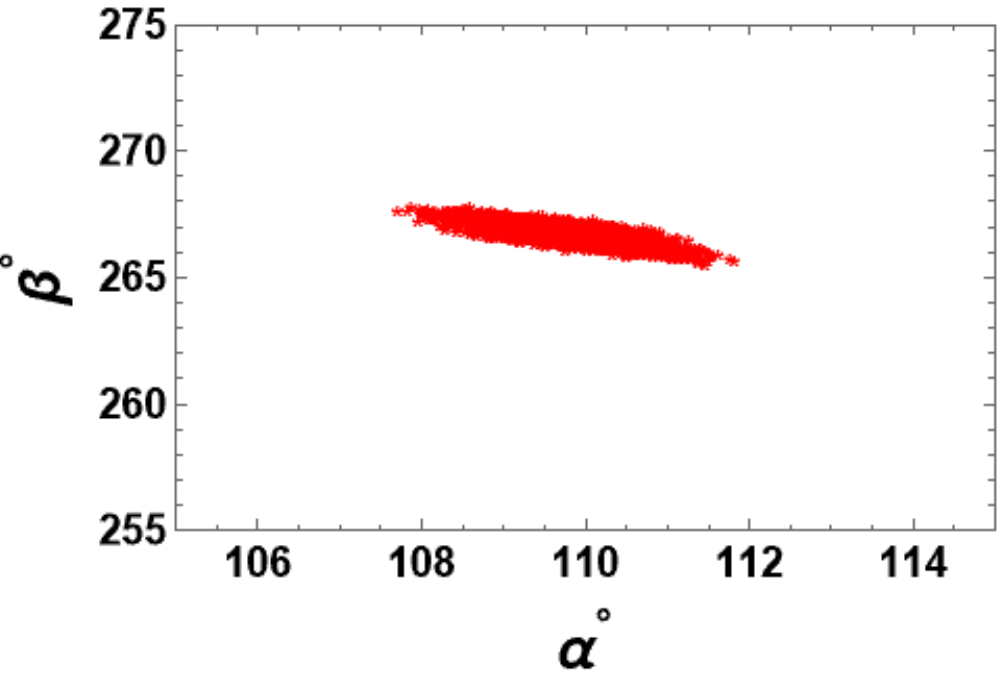}\label{fig:3(c)}}
    \subfigure[]{\includegraphics[width=0.25\textwidth]{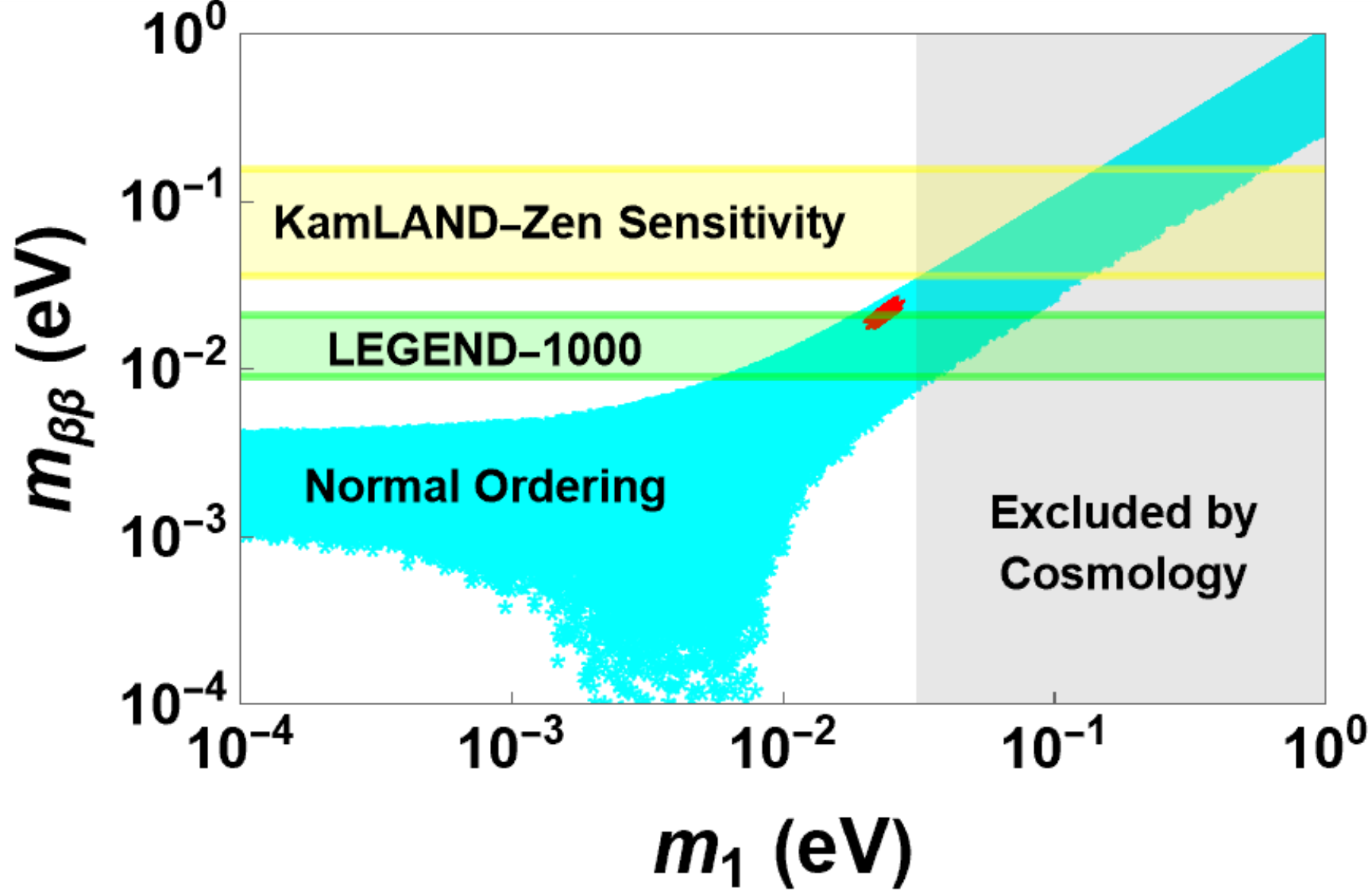}\label{fig:3(d)}}
    \caption{The correlation plots between (a) $\theta_{12}$ and $\theta_{13}$, (b) $\theta_{23}$ and $\delta$, (c)  $\alpha$ and $\beta$ for the texture $M_{\nu}^2$ under normal ordering of neutrino masses, (d) Shows the variation of $m_{\beta\beta}$ against the lightest neutrino mass ($m_1$)}
\label{fig:3}
\end{figure*}

\begin{figure}
  \centering
    \subfigure[]{\includegraphics[width=0.24\textwidth]{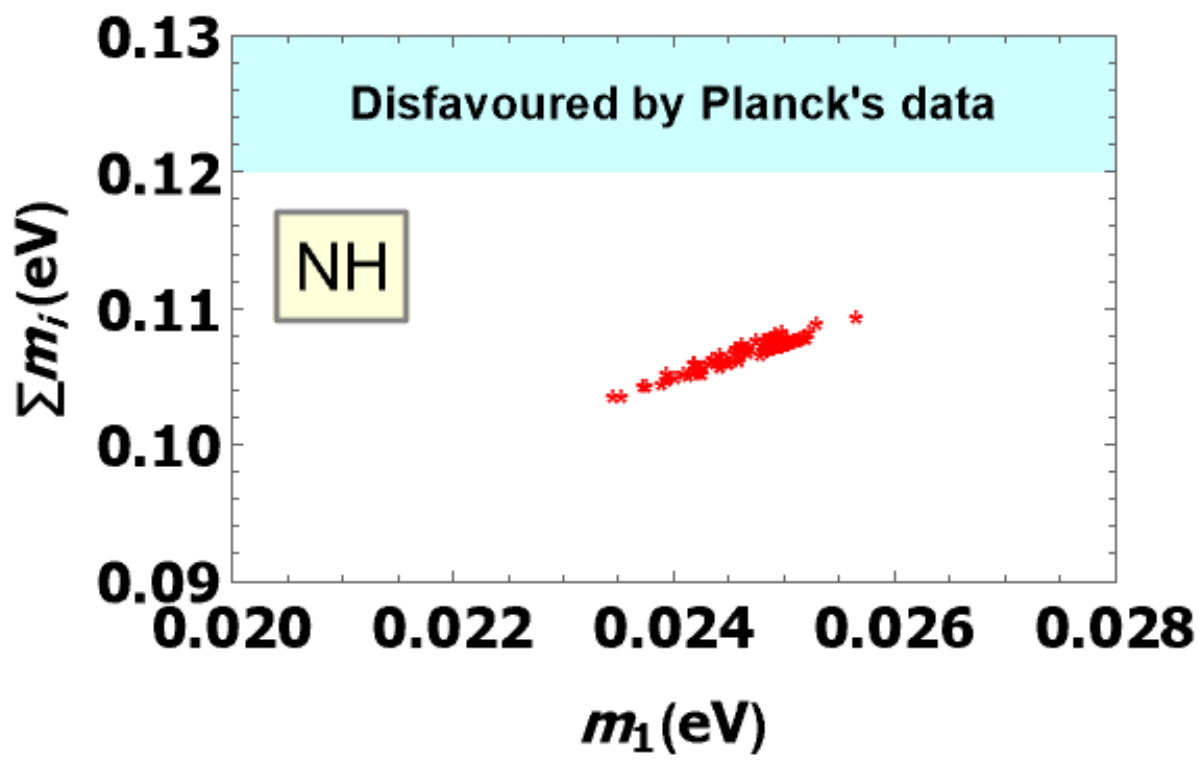}} 
    \subfigure[]{\includegraphics[width=0.24\textwidth]{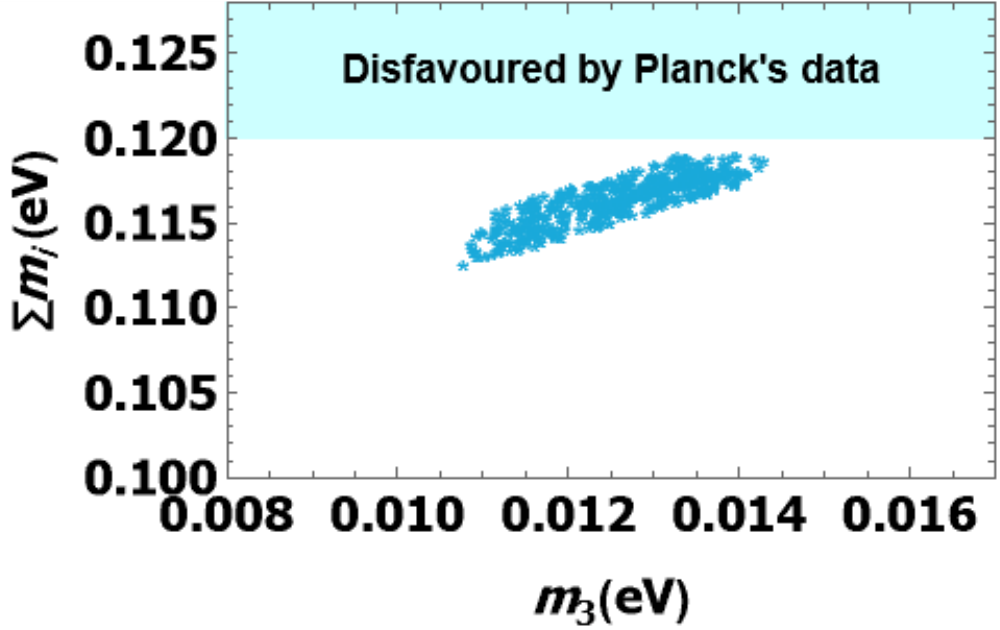}}
    \caption{The correlation plots between (a) $\Sigma m_i$ and $m_1$ for normal ordering, (b) $\Sigma m_i$ and $m_3$ for inverted ordering.}
\label{fig:sum}
\end{figure}

For $M^2_\nu$ as well, there is a corresponding dependent relation: $(M^2_{\nu})_{33}=-2(M^2_{\nu})_{12}$. This relation can be derived from the aforementioned three relations. Following the similar analysis, we generate the correlations plots for the parameters $\theta_{12}$, $\theta_{13}$, $\theta_{23}$, $\delta$, $\alpha$ and $\beta$ considering both normal and inverted orderings\,(see Figs.\,\ref{fig:3}). It is evident that for normal ordering, $\theta_{23}$ lies in the lower octant and a constrained bound for $\delta$ is obtained. In addition, the Majorana phases are constrained. For inverted ordering, the texture $M^2_\nu$ is ruled out as the prediction of $\theta_{13}$ lies outside the $3\sigma$ bound. For the above analysis, we set the two mass squared differences $\Delta m_{21}^2$ and $\Delta m_{31}^2$ at their respective $3\sigma$ bound\,\cite{Esteban:2024eli}. We estimate three mass eigenvalues $m_1$, $m_2$ and $m_3$ from two mass differences $\Delta m_{21}^2$ and $\Delta m_{31}^2$, such that $\Sigma m_i$ is consistent with the cosmological data\,\cite{Planck:2018vyg}. The variation of the sum of the three neutrino masses $\Sigma m_i$ with respect to the lightest neutrino mass for both orderings is shown in Fig.\,(\ref{fig:sum}). The unphysical phases $\phi_i$ being arbitrary can take any values between $0^\circ-360^\circ$.

\begin{figure*}
  \centering
    \subfigure[]{\includegraphics[width=0.24\textwidth]{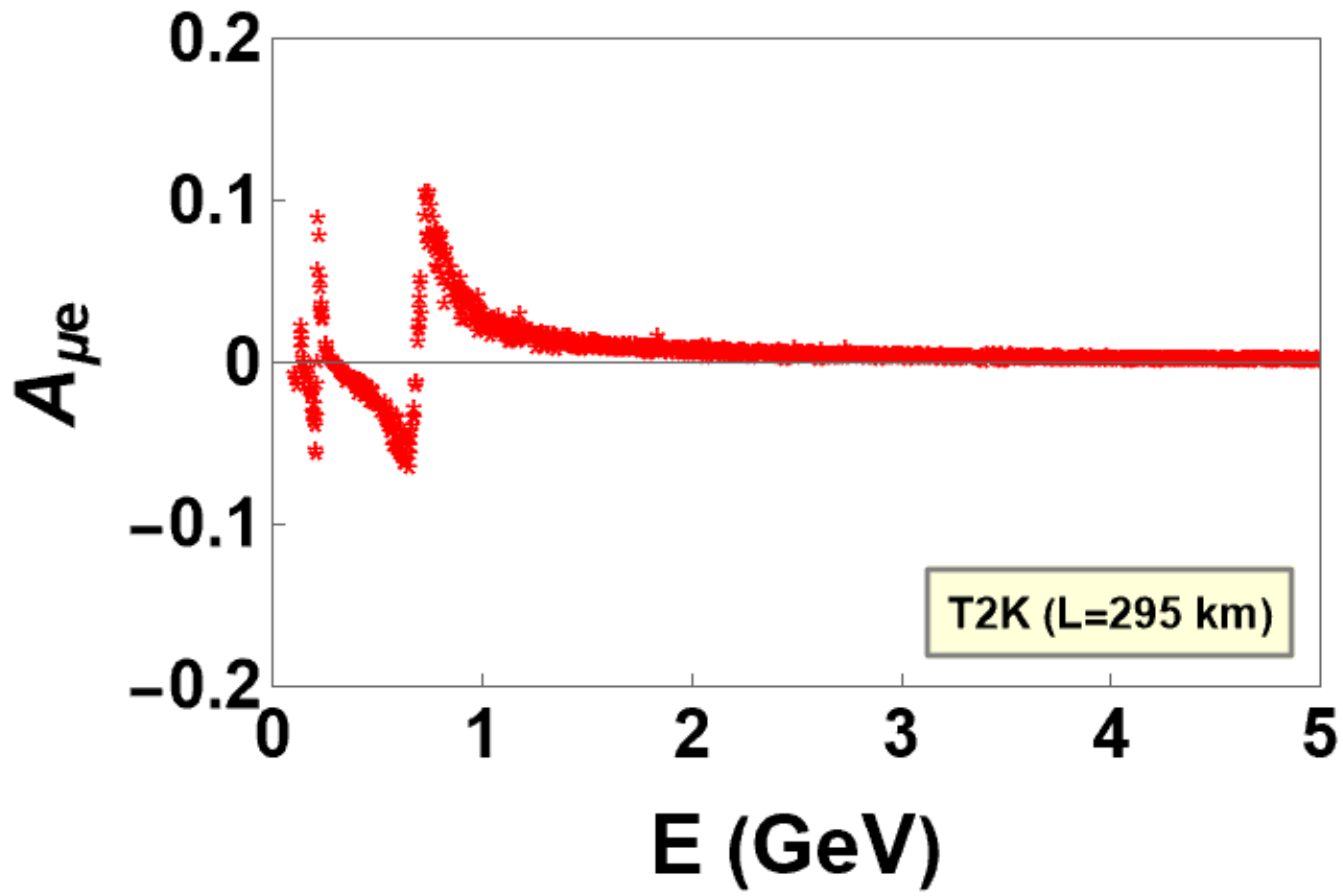}\label{fig:6(a)}} 
    \subfigure[]{\includegraphics[width=0.24\textwidth]{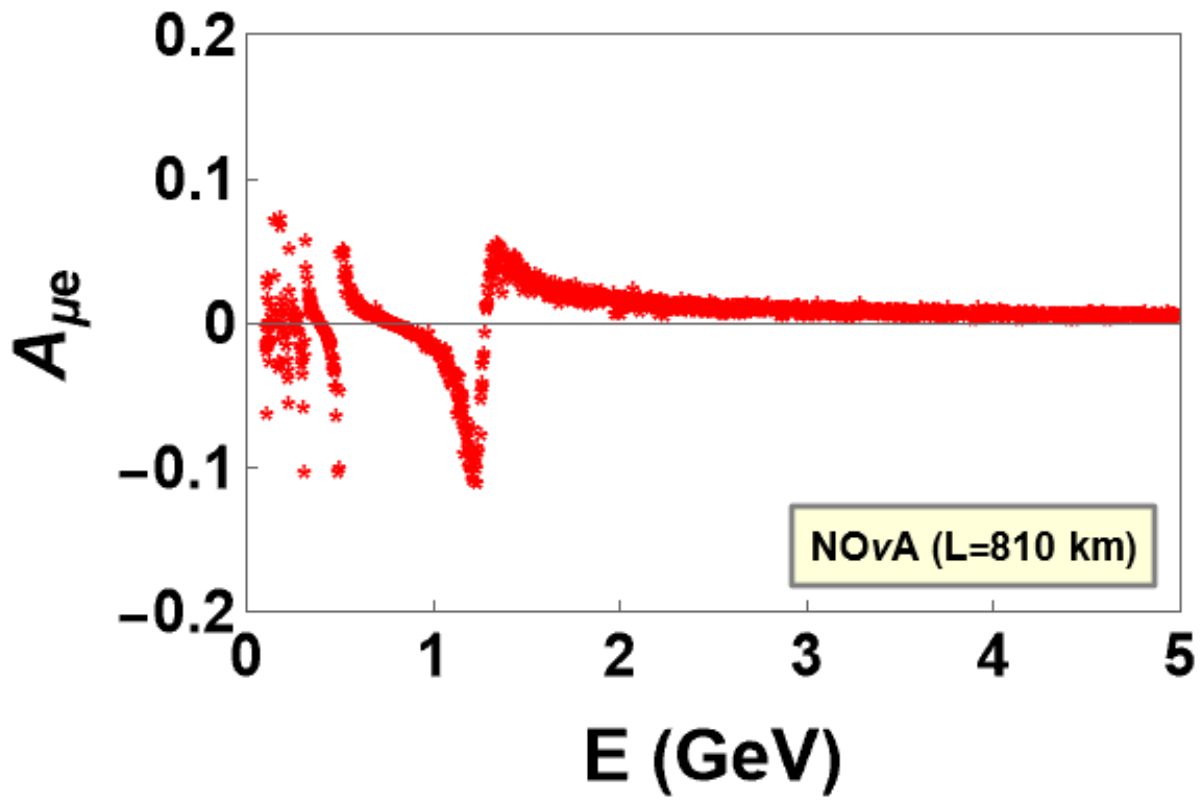}\label{fig:6(b)}} 
    \subfigure[]{\includegraphics[width=0.24\textwidth]{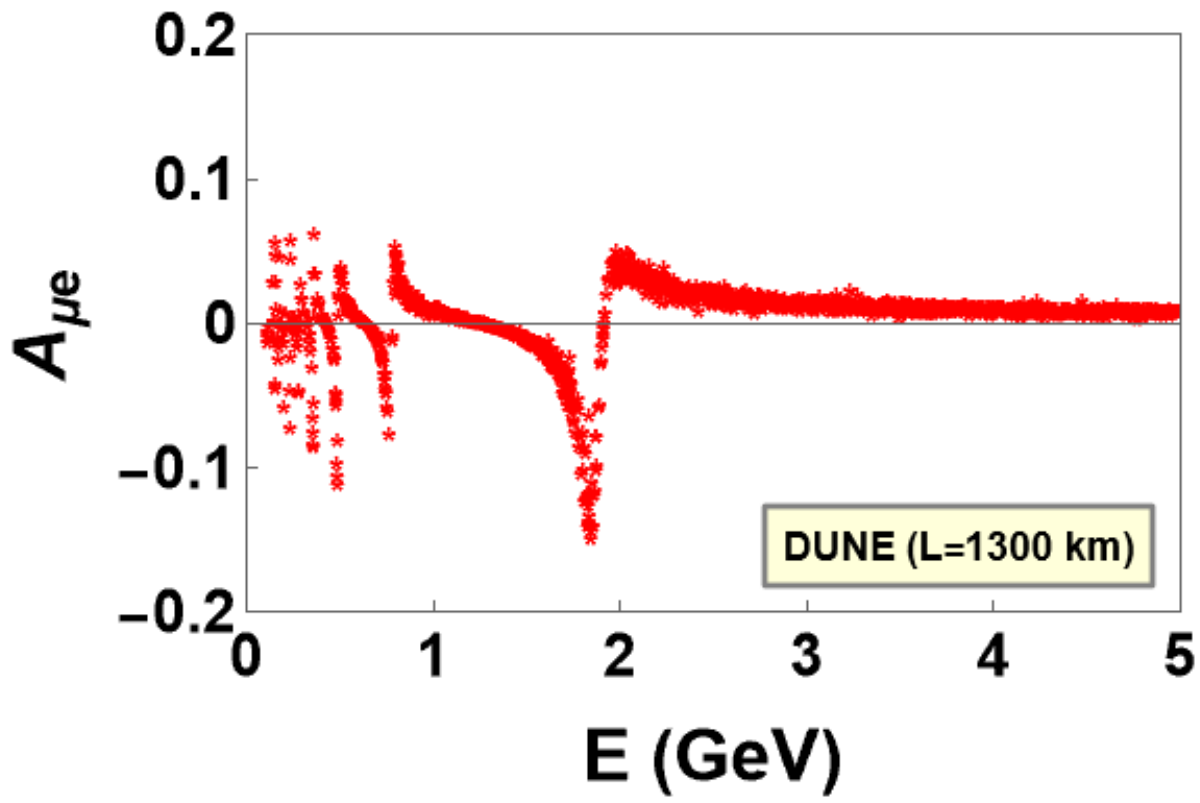}\label{fig:6(c)}}
    \caption{The correlation plots between (a) $A_{\mu e}$ vs $E$ for T2K $(L=295)$ km, (b) $A_{\mu e}$ vs $E$ for NO$\nu$A $(L=810)$ km, (c) $A_{\mu e}$ vs $E$ for DUNE $(L=1300)$ km for the texture $M_{\nu}^1$ under normal ordering of neutrino masses. }
\label{fig:6}
\end{figure*}

\subsection{\textbf{Effective Majorana Neutrino Mass}}

The justification of the Majorana nature of neutrinos is subjected to the discovery of neutrinoless double beta decay \cite{Schechter:1981bd}. The effective Majorana neutrino mass, denoted as $m_{\beta\beta}=|\sum_{k=1}^{3}{U^2_{1k}m_k}|$, is a observational parameter in neutrinoless double beta decay, where, $m_1$, $m_2$ and $m_3$ are the three mass eigenvalues, $U_{11}$,$U_{12}$ and $U_{13}$ are the elements of the PMNS matrix that carry the information of the Majorana phases. In recent years, several experiments have provided the upper bounds of $m_{\beta \beta}$: SuperNEMO(Se$^{82}$) as $67-131$ meV,  GERDA(Ge$^{76}$) as $104-228$ meV, EXO-200(Xe$^{136}$) as $111-477$ meV, CUORE(Te$^{130}$) as $75-350$ meV and KamLAND-Zen(Xe$^{136}$) as $61-165$ meV ~\cite{ Ejiri:2020xmm, Agostini:2022zub, CUORE:2019yfd, GERDA:2019ivs, KamLAND-Zen:2016pfg, SuperNEMO:2021hqx, CUORE:2018ncg, LEGEND:2021bnm}. In this regard, we predict the parameter $m_{\beta\beta}$ from the proposed textures\,(see Figs.\,\ref{fig:1(d)}, \ref{fig:2(d)}, \ref{fig:3(d)}). It is seen that for normal ordering, the prediction of $m_{\beta\beta}$ from the prosed textures lies partially in the sensitivity of the experiment LEGEND-$1000$\cite{LEGEND:2021bnm}. On the other hand, for inverted ordering, the same lies in the sensitivity of KamLAND-Zen \cite{KamLAND-Zen:2016pfg}.

\begin{figure*}
  \centering
    \subfigure[]{\includegraphics[width=0.24\textwidth]{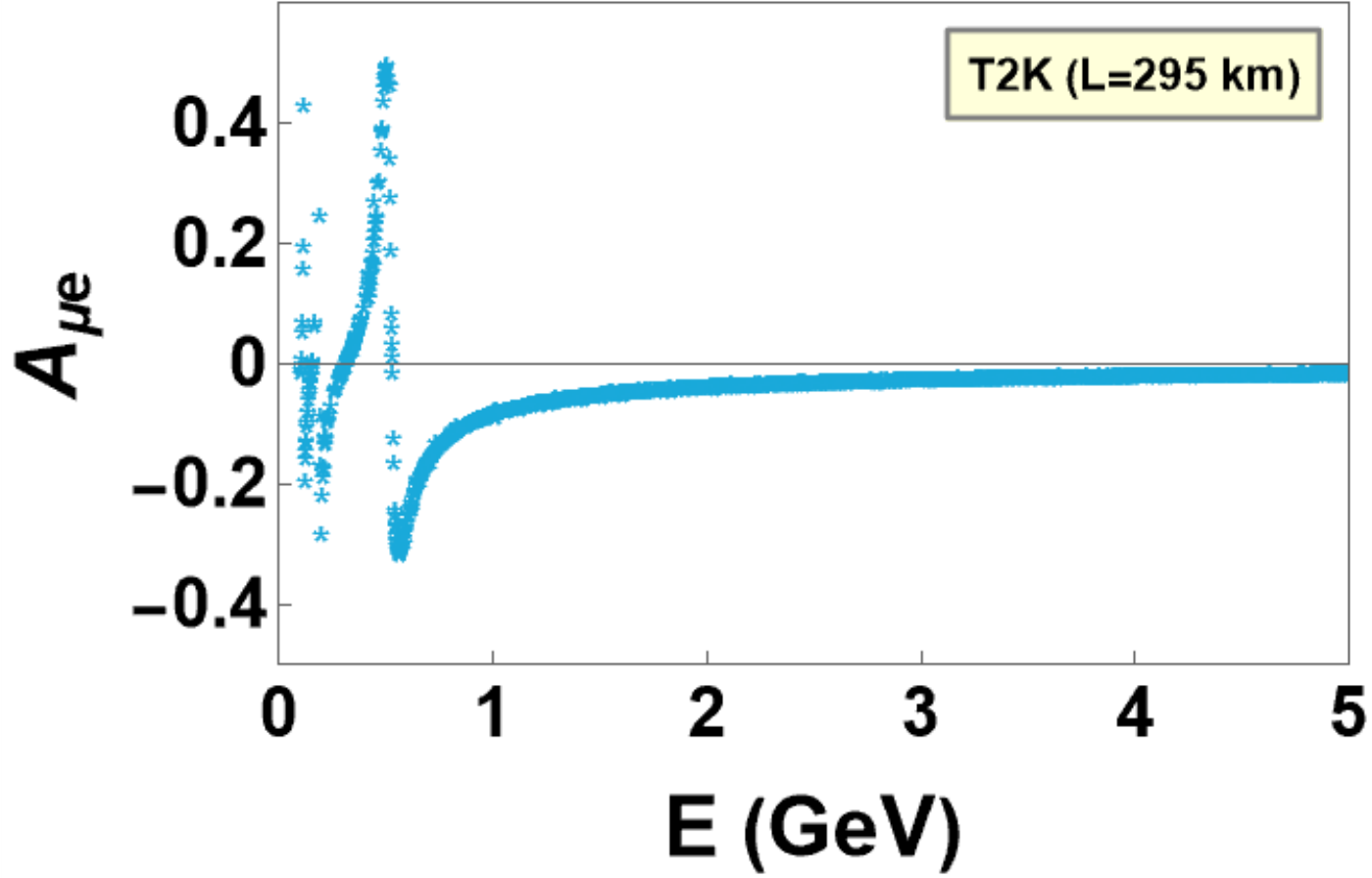}\label{fig:7(a)}} 
    \subfigure[]{\includegraphics[width=0.24\textwidth]{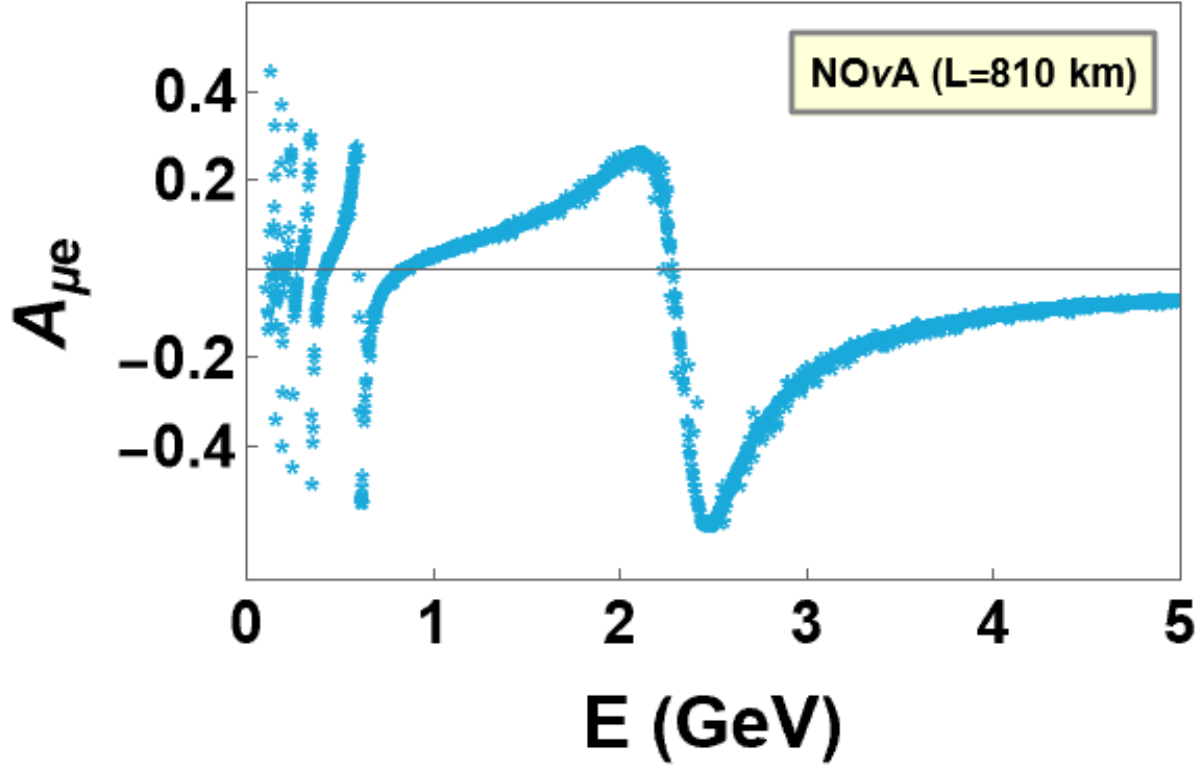}\label{fig:7(b)}} 
    \subfigure[]{\includegraphics[width=0.24\textwidth]{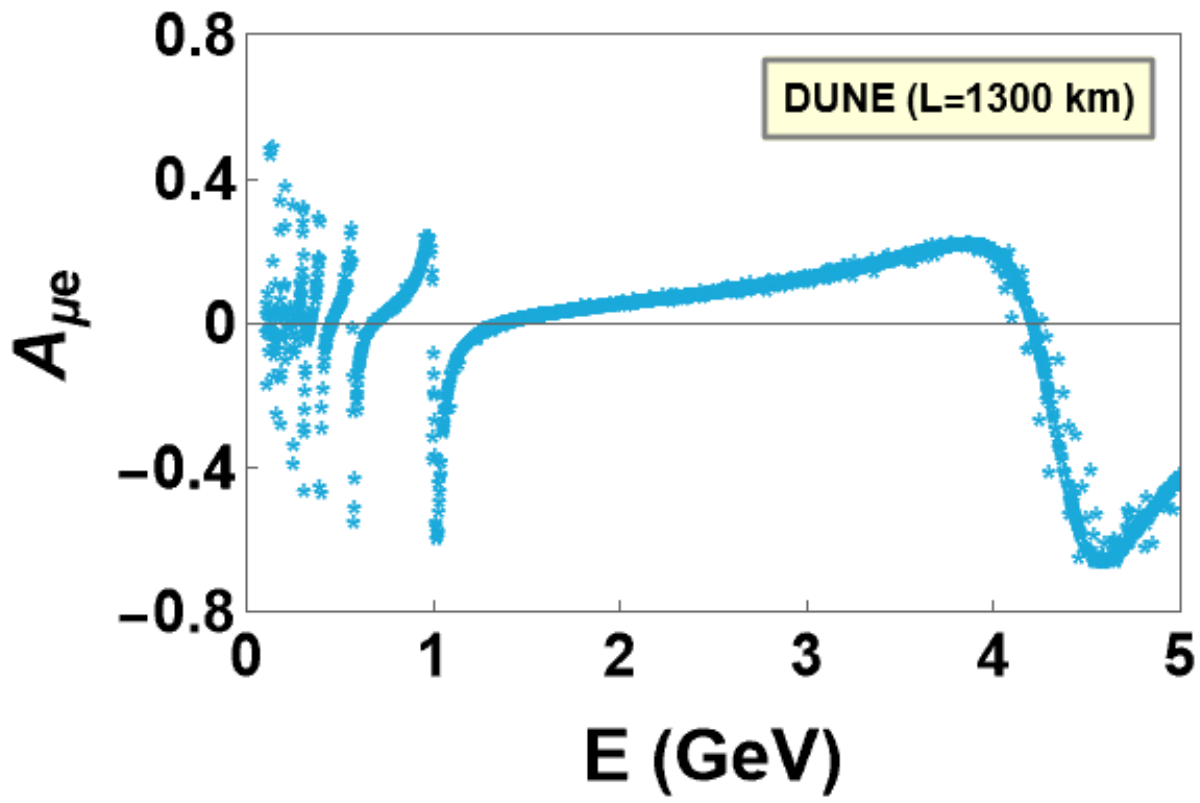}\label{fig:7(c)}}
    \caption{The correlation plots between (a) $A_{\mu e}$ vs $E$ for T2K $(L=295)$ km, (b) $A_{\mu e}$ vs $E$ for NO$\nu$A $(L=810)$ km, (c) $A_{\mu e}$ vs $E$ for DUNE $(L=1300)$ km for the texture $M_{\nu}^1$ under inverted ordering of neutrino masses.}
\label{fig:7}
\end{figure*}

\subsection{\textbf{CP Asymmetry Parameter}}

We study the effect of leptonic Dirac CP violation $\delta$ on neutrino oscillation through CP asymmetry parameter $(A_{\mu e})$. The said parameter is expressed as shown below,

\begin{equation}
A_{\mu e}=\frac{P(\nu_\mu \rightarrow \nu_e)-P(\bar{\nu}_\mu \rightarrow \bar{\nu}_e)}{P(\nu_\mu \rightarrow \nu_e)+P(\bar{\nu}_\mu \rightarrow \bar{\nu}_e)}.\nonumber
\end{equation}

In terms of the physical parameters, the transition probability $P(\nu_\mu \rightarrow \nu_e)$\,\cite{Sinha:2018xof} can be can be expressed as shown below,  

\begin{equation}
P(\nu_\mu \rightarrow \nu_e)= P_{\text{atm}}+P_{\text{sol}}+2 \sqrt{P_{\text{atm}}} \sqrt{P_{\text{sol}}} \cos(\Delta_{ij}+\delta),\nonumber
\end{equation}

where, 
$\Delta_{ij}=\Delta m_{ij}^2 \frac{L}{4 E}$. The $E$ stands for neutrino beam energy and $L$ represents the baseline length. The quantities $P_{\text{atm}}$ and $P_{\text{sol}}$ can be expressed as shown in the following,

\begin{eqnarray}
P_{\text{atm}}&=& \sin \theta_{23} \sin 2\theta_{13} \frac{\sin(\Delta_{31}-a L)}{(\Delta_{31}-a L)} \Delta_{31},\nonumber\\
P_{\text{sol}}&=& \cos \theta_{23} \sin 2\theta_{12} \frac{\sin(a L)}{(a L)} \Delta_{21}.\nonumber
\end{eqnarray}

The parameter $a=G_F N_e/\sqrt{2}$ depends on the neutrino propagation medium and occurs due to matter effects in neutrino propagation through the Earth. The factor $G_F$ represents  the Fermi constant and $N_e$ stands for the electron number density of the medium. For the earth, $a$ can be estimated as $3500{\rm km}^{-1}$\cite{Sinha:2018xof}. The CP asymmetry parameter can be expressed in terms of the physical parameters as shown in the following,

\begin{equation}
A_{\mu e}=
\frac{2\sqrt{P_{\text{atm}}}\sqrt{P_{\text{sol}}}\sin\Delta_{32}\sin\delta}{P_{\text{atm}}+2\sqrt{P_{\text{atm}}}\sqrt{P_{\text{sol}}}\cos\Delta_{32}\cos\delta+P_{\text{sol}}}.\nonumber
\end{equation}

In the present work, we visualize the variation of CP asymmetry parameter for three experiments with different baseline lengths: T$2$K\,($L=295$ km), No$\nu$A\,($L=810$ km) and DUNE\,($L=1300$ km) by considering both normal and inverted orderings (see Figs. \ref{fig:6}-\ref{fig:7}). For the above analysis, the beam energy $E$ is set at $(0.5-5)$\,GeV. We wish to highlight that we visualize the variation of the CP asymmetry parameter based on texture prediction by adhering to different baseline lengths and beam energies. The $A_{\mu e}$ largely depends on the Dirac CP phase $\delta$. The long-baseline experiments aim to measure the said phase by taking into consideration the sub-dominant matter effects. The results of T$2$K ($L=295$ km) and No$\nu$A ($L=810$ km) on the prediction of $\delta$ differ for the inverted ordering. Future experiments such as DUNE, with $L=1300$ km and a peak beam energy around $2.5$ GeV, may shed light on this discrepancy, and our prediction of $A_{\mu e}$ can be tested.

		\begin{table*}
\centering
\begin{tabular*}{\textwidth}{@{\extracolsep{\fill}} cccccccc} 
\hline \hline
Fields & $D_{l_{L}}$ & $l_{R}$ & $\nu_{l_R}$ & $H$ & $\Phi$ & $\zeta$ & $\Delta$\\ 
			\hline\hline
			$SU(2)_{L}$ & 2 & 1 & 1 & 2 & 2 & 1 & 3 \\
			\hline
			$U(1)_{Y}$ & -1 & -2 & 0 & 1 & -1 & -2 & 3 \\
			\hline
			$A_{4}$ & 3 & (1,\,1{$'$},\,1{$''$}) & (1,\,1{$'$},\,1{$''$}) & 3 & 3 & 1 & 3 \\
			\hline
			$Z_3$ & 1 & ($\omega^2$,\,$\omega^2$,\,$\omega^2$) & ($\omega$,\,$\omega$,\,$\omega$) & $\omega$ & $\omega^2$ & $\omega$ & 1 \\
			\hline
			\end{tabular*}
		\caption{The transformation properties of various fields under $SU(2)_{L} \times A_{4} \times Z_3$.} 
		\label{Field Content of M1}
	\end{table*}

\begin{figure*}
  \centering
    \subfigure[]{\includegraphics[width=0.24\textwidth]{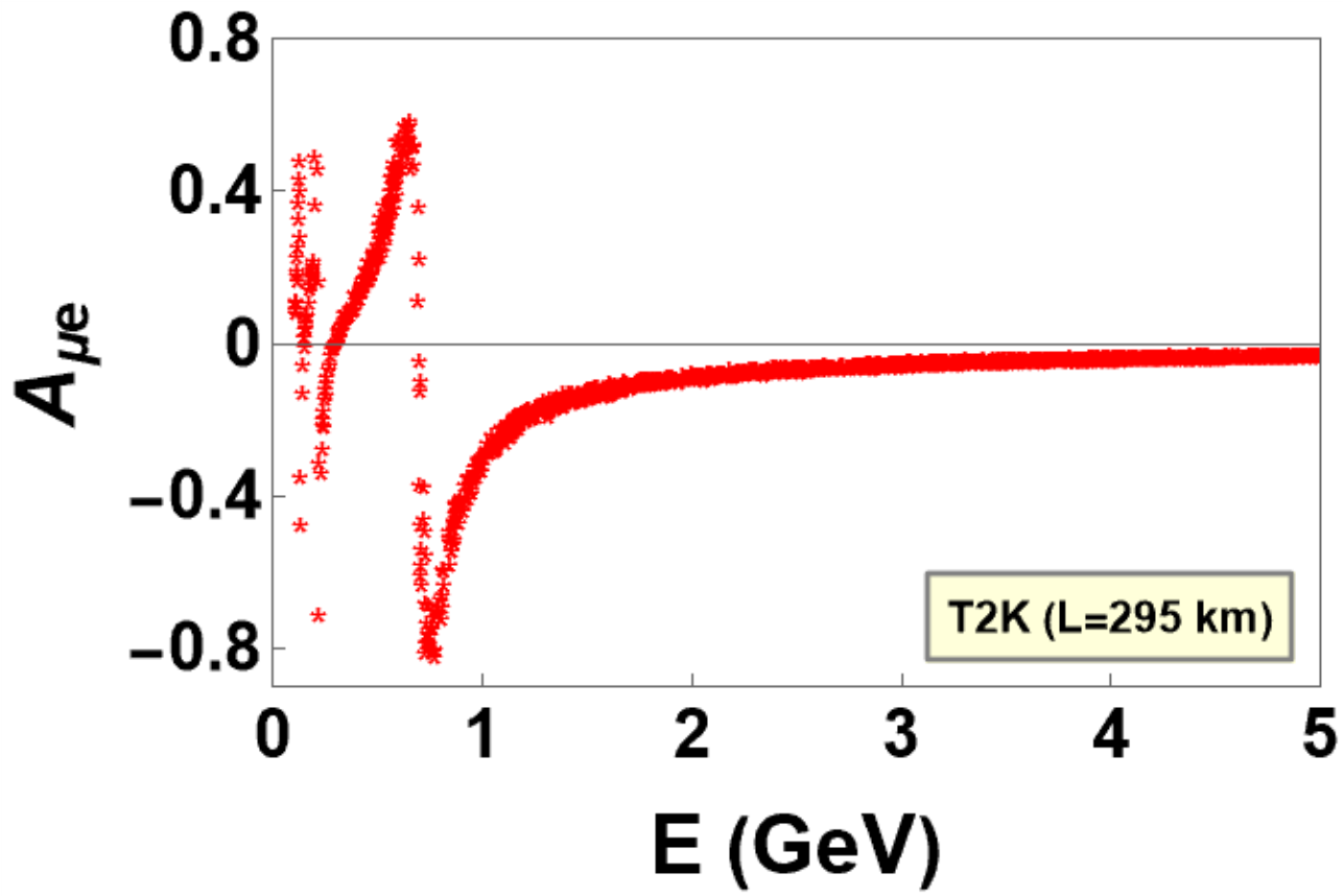}\label{fig:8(a)}} 
    \subfigure[]{\includegraphics[width=0.24\textwidth]{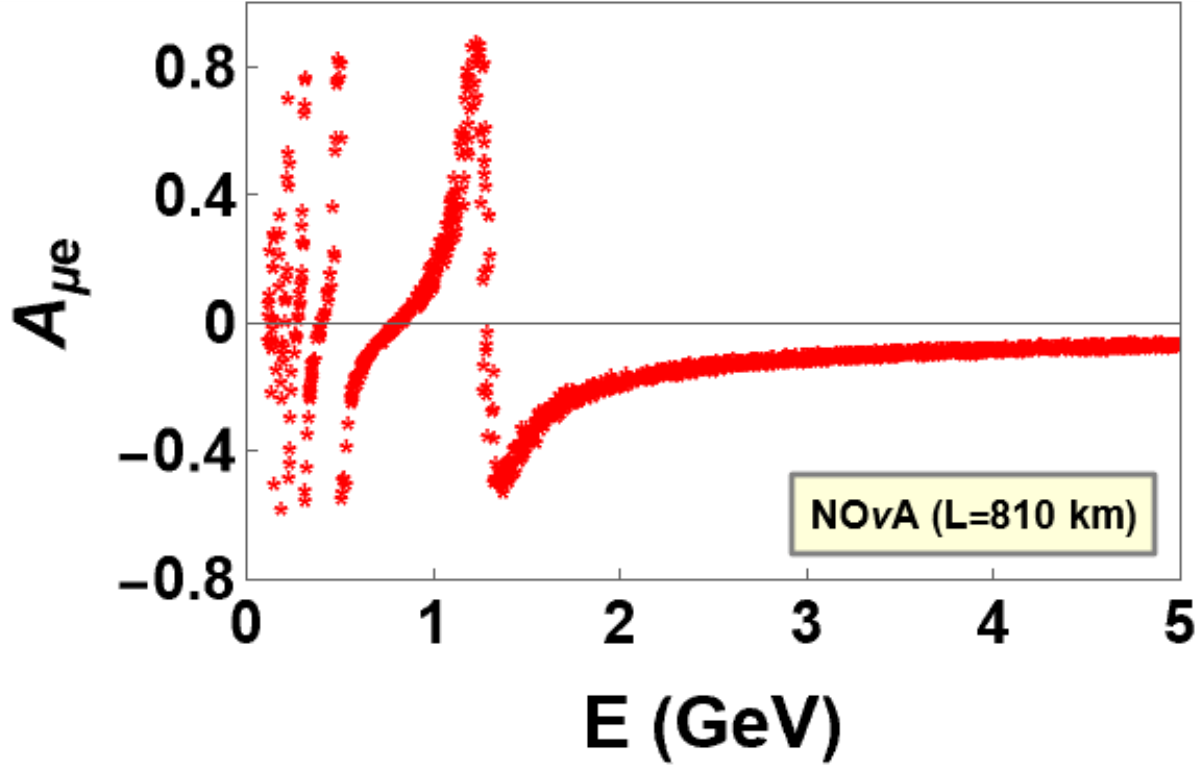}\label{fig:8(b)}} 
    \subfigure[]{\includegraphics[width=0.24\textwidth]{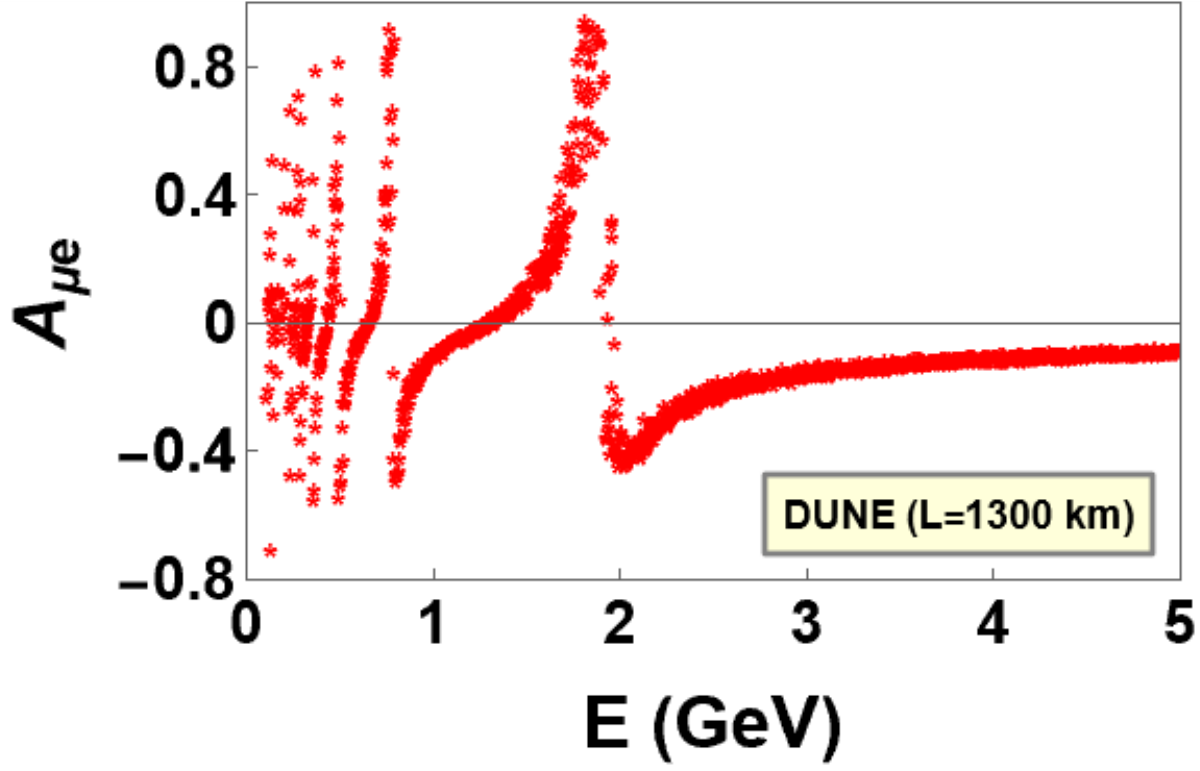}\label{fig:8(c)}}
    \caption{The correlation plots between (a) $A_{\mu e}$ vs $E$ for T2K $(L=295)$ km, (b) $A_{\mu e}$ vs $E$ for NO$\nu$A $(L=810)$ km, (c) $A_{\mu e}$ vs $E$ for DUNE $(L=1300)$ km for the texture $M_{\nu}^2$ under normal ordering of neutrino masses. }
\label{fig:8}
\end{figure*}

\section{Symmetry Realization}

To understand the proposed texture from symmetry framework, we try to incorporate see-saw mechanism in association with $A_4$ group. In this regard, we extend the field content of the SM by introducing three right-handed neutrinos ($\nu_{e_R}$, $\nu_{\mu_R}$, $\nu_{\tau_R}$), a scalar singlet ($\zeta$), a scalar doublet ($\phi$) and a scalar triplet ($\Delta$). The transformation properties of the field content are summarised in Table\,\ref{Field Content of M1}. The $SU(2)_L \times A_4 \times Z_3$ invariant Lagrangian is constructed in the following way,

\begin{eqnarray}
- \mathcal{L}_Y &=& y_{e}(\bar{D}_{l_{L}}H)_{1}e_{R_{1}} + y_{\mu}(\bar{D}_{l_{L}}H)_{1'}\mu_{R_{1''}} + y_{\tau}(\bar{D}_{l_{L}}H)_{1''}\nonumber\\&&\tau_{R_{1'}}+ y_{1}(\bar{D}_{l_{L}}\Phi)_{1}\nu_{eR_{1}}+ y_{2}(\bar{D}_{l_{L}}\Phi)_{1''}\nu_{\mu R_{1'}} + y_{3}\nonumber\\&& (\bar{D}_{l_{L}}\Phi)_{1'}\nu_{\tau_{1''}}+ \frac{1}{2}y_{R_{1}}(\overline{\nu^c}_{e_{R}}\nu_{e_{R}})_{1}\zeta
		+ \frac{1}{2}y_{R_{2}}[(\overline{\nu^c}_{\mu_{R}}\nonumber\\&&\nu_{\tau_{R}})  + (\overline{\nu^c}_{\tau_{R}}\nu_{\mu_{R}})]\zeta+ y_{T_{2}}(\bar{D}_{l_{L}}D_{l_{L}}^{c})_{3_{S}}\Delta_{3} + h.c.\nonumber
\label{Yukawa Lagrangian M1}
\end{eqnarray}

 The additional group $Z_3$ is incorporated to cut short some undesired terms that are allowed by $A_4$. The vacuum expectation values\,(vev) for the scalar fields are chosen as: $\langle H \rangle_{0}=v_{H}(1,1,1)^{T}$ and $\langle\Phi\rangle_{0}=v_{\Phi}(1,1,1)^{T}, \langle\zeta\rangle_{0}=v_{\zeta}$. 
 
 Incorporating the vevs as discussed above, we derive the charged lepton mass matrix($M_{L}$) in the following form,

	\begin{equation}
		M_{l}=v_{H} \begin{bmatrix}
			y_{e} & y_{\mu} & y_{\tau} \\
			y_{e} & \omega\,y_{\mu} & \omega^2\, y_{\tau} \\
			y_{e} & \omega^2\, y_{\mu} & \omega \,y_{\tau} \\
		\end{bmatrix}, \nonumber
	\end{equation}

	We diagonalize $M_l$ as $M_{l}^{diag}=U_{l_{L}}^{\dagger}M_{l}U_{l_R}$, where $M^{diag}_{l}=\sqrt{3}\,diag(y_{e},y_{\mu},y_{\tau})$. The $U_{l_{L}}$ and $U_{l_{R}}$ are expressed as,
	\begin{eqnarray}
	U_{l_{L}}&=&\frac{1}{\sqrt{3}} \begin{bmatrix}
			e^{i\eta} &  e^{i\kappa} & e^{i\xi}\\
			e^{i\eta} & \omega \,e^{i\kappa} & \omega^2\, e^{i\xi}\\
			e^{i\eta} & \omega^2\,e^{i\kappa} & \omega\,e^{i\xi}\\
		\end{bmatrix}\\
		U_{l_{R}}&=&diag\begin{bmatrix}
			
		e^{i\eta}, & e^{i\kappa}, & \,e^{i\xi}
		\end{bmatrix}, \nonumber
	\end{eqnarray}

	 The arbitrary phases $\eta$, $\kappa$ and $\xi$ are added in $U_{l_{L}}$ and $U_{l_{R}}$ with the motivation that the choice of eigenvectors is not unique.
	
	The Dirac neutrino mass matrix($M_{D}$) and the right-handed neutrino mass matrix($M_{R}$) are obtained as shown in the following,
	
	\begin{equation}
		M_{D}=v_{\Phi} \begin{bmatrix}
			y_1 & y_2 & y_3 \\
			y_1 & \omega\,y_2 & \omega^{2}\,y_3 \\
			y_1 & \omega^{2}\,y_2 & \omega\,y_3 \\
		\end{bmatrix},
	\quad \quad
		M_{R}= \begin{bmatrix}
			P & 0 & 0 \\
			0 & 0 & Q \\
			0 & Q & 0\\
		\end{bmatrix}, \nonumber
	\end{equation}
	
	where, $\omega=e^{2i\pi/3}$, $P= \frac{1}{2} y_{R_{1}}v_{\zeta}$ and $Q= \frac{1}{2} y_{R_{2}}v_{\zeta}$.
	
 The Type-I seesaw contribution to the neutrino mass matrix is taken as $M_{T_{1}}= -M_{D}M_{R}^{-1}M_{D}^{T}$. The Type-II seesaw contribution with the specific choice of vev $\langle\Delta\rangle_{0}=v_{\Delta}(0,1,0)^T$ comes as shown below,
	\begin{equation}
		M_{T_{2}}=y_{T_{2}}v_{\Delta} \begin{bmatrix}
			0 & 0 & 1\\
			0 & 0 & 0 \\
			1 & 0  & 0\\
		\end{bmatrix}. \nonumber
	\end{equation}

The neutrino mass matrix is constructed by taking the contribution from Type-I and Type-II seesaw mechanisms. Now, to move to a basis where the charged lepton mass matrix is diagonal, we incorporate a certain transformation $(U_{l_{L}}^{T}M_{\nu_{s}}U_{l_{L}})$ and obtain the effective neutrino mass matrix as shown below,  

\begin{eqnarray}
M^1_{\nu}&=&\begin{bmatrix}
A &  B  &   (i B)^* \\
B &  2B^*  &  D \\
(i B)^* & D &  -2B \\
\end{bmatrix}.
\end{eqnarray}

 It is to be highlighted that to obtain the texture $M^1_{\nu}$, we choose $\eta=\frac{\pi}{2}$, $\kappa=\frac{\pi}{6}$ and $\xi=\frac{\pi}{3}$.
	
The proposed texture $M^2_\nu$ can be derived from the same Lagrangian appearing in Eq.\,(\ref{Yukawa Lagrangian M1}) under the specific choice of vev $\langle\Delta\rangle_{0}=v_{\Delta}(\omega^2,0,1)^T$ and by setting $\eta=\frac{\pi}{6}$, $\kappa=\frac{3 \pi}{6}$ and $\xi=\frac{2 \pi}{6}$. The above mass matrices derived from the symmetry framework are expressed in terms of model parameters involving several vevs and Yukawa couplings resembles the exact form of two proposed neutrino mass matrix texture as given in Eq.\,(\ref{M1}).

\section{Summary}

The novel $\mu-\tau$ symmetry is well established in the literature, with the prediction of maximal $\theta_{23}$ being inherent. However, since it is ruled out by experiments due to its prediction of $\theta_{13} = 0$, several attempts have been made to deviate from exact $\mu-\tau$ symmetry. In the present work, we introduce new approaches to break $\mu-\tau$ symmetry. In this regard, we propose two new $\mu-\tau$-deviated neutrino mass matrix textures, $M_\nu^1$ and $M_\nu^2$, and study their phenomenological implications. The mixing schemes provided by both textures are consistent with experimental observations. Notably, both textures predict $\theta_{23}$ in the lower octant and impose sharp constraints on $\delta$ for the normal ordering of neutrino masses. For the inverted ordering, $M_\nu^1$ remains consistent with experimental observations, whereas $M_\nu^2$ is ruled out. As a potential application, we obtain predictions for the parameter $m_{\beta\beta}$ based on the proposed textures. Additionally, we visualize the variation of the CP asymmetry parameter based on texture predictions for three different baseline lengths. To establish the theoretical foundation of the proposed textures, we realize them within the framework of the seesaw mechanism, starting from the $SU(2)_L \times A_4 \times Z_3$ symmetry group.

We wish to mention that the model sector can be extended to study phenomena such as charged lepton flavour violation,(cLFV). In the SM, the cLFV process is very rare, with a branching ratio around $10^{-50}$. Among various decay channels, the dominant one is $\mu\rightarrow e \gamma$. The current experiment MEG has set an upper limit on the branching ratio: BR$(\mu\rightarrow e \gamma)<4.2\times 10^{-13}(90\%,\text{CL})$\cite{MEG:2016leq}. However, in the present work, we focus mainly on the texture sector, and this part of the study is beyond the scope of the present work.

\section{Acknowledgement}

The research work of PC is supported by Innovation in Science Pursuit for Inspired Research (INSPIRE), Department of Science and Technology, Government of India, New Delhi vide grant No. IF190651.

\section{The Scalar Potential}

The $SU(2)_L \times A_4 \times Z_3$ invariant scalar potential can be constructed as shown in the following,

\scriptsize
\begin{eqnarray}
V &=& V(H)+V(\Phi)+V(\Delta)+V(\zeta)+V(H, \Phi)+\nonumber\\&&V(H,\Delta)+V(H,\zeta)+V(\Phi,\Delta)+V(\Phi,\zeta)+V(\Delta,\zeta),\nonumber
\end{eqnarray}

where,

\begin{eqnarray}
V(H)&=& -\mu^2_H (H^\dagger H)+\lambda^H_1 (H^\dagger H) (H^\dagger H)+\lambda^H_2 (H^\dagger H)_{1'} \nonumber\\&&(H^\dagger H)_{1''}+\lambda^H_3 (H^\dagger  H)_{3_s} (H^\dagger H)_{3_s}+\lambda^H_4 (H^\dagger\nonumber\\&& H)_{3_s} (H^\dagger H)_{3_a}+\lambda^H_5 (H^\dagger H)_{3_a} (H^\dagger H)_{3_a},\nonumber\\
V(\zeta)&=& -\mu^2_\zeta (\zeta^\dagger \zeta)+\lambda^H (\zeta^\dagger \zeta)(\zeta^\dagger \zeta),\nonumber\\
V(\Phi)&=& -\mu^2_\Phi (\Phi^\dagger \Phi)+\lambda^\Phi_1 (\Phi^\dagger \Phi) (\Phi^\dagger \Phi)+\lambda^\Phi_2 (\Phi^\dagger \Phi)_{1'} (\Phi^\dagger\nonumber\\&& \Phi)_{1''}+\lambda^\Phi_3 (\Phi^\dagger \Phi)_{3_s} (\Phi^\dagger \Phi)_{3_s}+\lambda^\Phi_4 (\Phi^\dagger \Phi)_{3_s} (\Phi^\dagger \nonumber\\&&\Phi)_{3_a}+\lambda^\Phi_5 (\Phi^\dagger \Phi)_{3_a} (\Phi^\dagger \Phi)_{3_a},\nonumber\\
V(H, \zeta)&=& \lambda^{H \zeta}_1 (H^\dagger H)(\zeta^\dagger \zeta),\nonumber\\
V(\Phi, \zeta)&=& \lambda^{\Phi \zeta}_1 (\Phi^\dagger \Phi)(\zeta^\dagger \zeta),\nonumber
\end{eqnarray}

\begin{eqnarray}
V(\Delta)&=& -\mu^2_\Delta Tr(\Delta^\dagger \Delta)+\lambda^\Delta_1 Tr(\Delta^\dagger \Delta) Tr(\Delta^\dagger \Delta)+\lambda^\Delta_2\nonumber\\&& Tr(\Delta^\dagger \Delta)_{1'} Tr(\Delta^\dagger \Delta)_{1''}+\lambda^\Delta_3 Tr(\Delta^\dagger \Delta)_{3_s} Tr(\nonumber\\&&\Delta^\dagger \Delta)_{3_s}+\lambda^\Delta_4Tr(\Delta^\dagger \Delta)_{3_s} Tr(\Delta^\dagger \Delta)_{3_a}+\lambda^\Delta_5 \nonumber\\&& Tr(\Delta^\dagger \Delta)_{3_a} Tr(\Delta^\dagger \Delta)_{3_a},\nonumber\\
V(\Delta, \zeta)&=& \lambda^{\Delta \zeta}_1 Tr(\Delta^\dagger \Delta)(\zeta^\dagger \zeta),\nonumber\\
V(H, \Phi )&=& \lambda^{H \Phi}_1 (H^\dagger H)(\Phi^\dagger \Phi)+\lambda^{H \Phi}_2 [(H^\dagger H)_{1'}(\Phi^\dagger \Phi)_{1''}\nonumber\\&&+(H^\dagger H)_{1''}(\Phi^\dagger \Phi)_{1'}]+\lambda^{H \Phi}_3  (H^\dagger H)_{3_s}(\Phi^\dagger \Phi)_{3_s}\nonumber\\&&+\lambda^{H \Phi}_4 [(H^\dagger H)_{3_s}(\Phi^\dagger \Phi)_{3_a}+(H^\dagger H)_{3_a}(\Phi^\dagger \nonumber\\&&\Phi)_{3_s}]+\lambda^{H \Phi}_5  (H^\dagger H)_{3_a}(\Phi^\dagger \Phi)_{3_a},\nonumber\\
V(H, \Delta )&=& \lambda^{H \Delta}_1 (H^\dagger H)Tr(\Delta^\dagger \Delta)+\lambda^{H \Delta}_2 [(H^\dagger H)_{1'}Tr(\Delta^\dagger\nonumber\\&& \Delta)_{1''}+(H^\dagger H)_{1''}Tr(\Delta^\dagger \Delta)_{1'}]+\lambda^{H \Delta}_3  (H^\dagger H)_{3_s}\nonumber\\&&Tr(\Delta^\dagger \Delta)_{3_s}+\lambda^{H \Delta}_4 [(H^\dagger H)_{3_s}Tr(\Delta^\dagger  \Delta)_{3_a}+\nonumber\\&&(H^\dagger H)_{3_a}Tr(\Delta^\dagger \Delta)_{3_s}]+\lambda^{H \Delta}_5  (H^\dagger H)_{3_a}Tr(\Delta^\dagger\nonumber\\&& \Delta)_{3_a}+\mu_1 (H^T i \sigma_2 \Delta^\dagger H),\nonumber\\
V(\Phi, \Delta )&=& \lambda^{\Phi \Delta}_1 (\Phi^\dagger \Phi)Tr(\Delta^\dagger \Delta)+\lambda^{\Phi \Delta}_2 [(\Phi^\dagger \Phi)_{1'}Tr(\Delta^\dagger\nonumber\\&& \Delta)_{1''}+(\Phi^\dagger \Phi)_{1''}Tr(\Delta^\dagger \Delta)_{1'}]+\lambda^{\Phi \Delta}_3  (\Phi^\dagger \Phi)_{3_s}\nonumber\\&&Tr(\Delta^\dagger \Delta)_{3_s}+\lambda^{\Phi \Delta}_4 [(\Phi^\dagger \Phi)_{3_s}Tr(\Delta^\dagger \Delta)_{3_a}+(\Phi^\dagger \nonumber\\&&\Phi)_{3_a}Tr(\Delta^\dagger \Delta)_{3_s}]+\lambda^{\Phi \Delta}_5  (\Phi^\dagger \Phi)_{3_a}Tr(\Delta^\dagger \Delta)_{3_a},\nonumber
\end{eqnarray}

 Based on the minimization of the scalar potential, the following equations hold good for the chosen vacuum expectation values (VEVs) $\langle H \rangle_{\circ}=v_{H}\,(1,1,1)^{T}$, $\langle \Phi \rangle_{\circ}=v_{\Phi}\,(1,1,1)^{T}$, $\langle \Delta \rangle_{\circ}=v_{\Delta}\,(0,1,0)^{T}$ and $\langle \zeta \rangle_{\circ}=v_{\zeta}$\,: 
 
 \begin{eqnarray}
\frac{\partial V}{\partial H_1}&=& v_H (-\mu _H^2+(\lambda _1^{H\Delta}-\lambda _2^{H\Delta }) v_{\Delta }^2+2 \mu_1 v_{\Delta }+\lambda _1^{H\zeta } v_{\zeta }^2+\nonumber\\&&\left(6 \lambda _1^H+8 \lambda _3^H\right) v_H^2+ 3 \lambda _1^{\Phi H} v_{\Phi }^2+4 \lambda _3^{\Phi H} v_{\Phi }^2)=0,\nonumber\\
\frac{\partial V}{\partial H_2}&=& v_H (-\mu _H^2+(6 \lambda _1^H+8 \lambda _3^H) v_H^2+(\lambda _1^{H\Delta}+2 \lambda _2^{H\Delta }) v_{\Delta }^2\nonumber\\&&+\lambda _1^{H\zeta } v_{\zeta }^2+3 \lambda _1^{\Phi H} v_{\Phi }^2+4 \lambda _3^{\Phi H} v_{\Phi }^2)=0,\nonumber\\
\frac{\partial V}{\partial H_3}&=& v_H (-\mu _H^2+(6 \lambda _1^H+8 \lambda _3^H) v_H^2+(\lambda _1^{H\Delta}-\lambda _2^{H\Delta }) v_{\Delta }^2\nonumber\\&&+\lambda _1^{H\zeta} v_{\zeta }^2+3 \lambda _1^{\Phi H} v_{\Phi }^2+4 \lambda _3^{\Phi H} v_{\Phi }^2)=0,\nonumber\\
\frac{\partial V}{\partial \Phi_1}&=& v_{\Phi } (3 v_H^2 \lambda _1^{\Phi H}+4 v_H^2 \lambda _3^{\Phi H}-\mu _{\Phi }^2+\lambda _1^{\Phi \Delta } v_{\Delta }^2-\lambda _2^{\Phi \Delta }\nonumber\\&& v_{\Delta }^2+\lambda _1^{\Phi \zeta } v_{\zeta }^2+6 \lambda _1^{\Phi } v_{\Phi }^2+8 \lambda _3^{\Phi } v_{\Phi }^2)=0,\nonumber\\
\frac{\partial V}{\partial \Phi_2}&=& v_{\Phi } (3 v_H^2 \lambda _1^{\Phi H}+4 v_H^2 \lambda _3^{\Phi H}-\mu _{\Phi }^2+\lambda _1^{\Phi \Delta } v_{\Delta }^2+\nonumber\\&&2 \lambda _2^{\Phi \Delta } v_{\Delta }^2+\lambda _1^{\Phi \zeta } v_{\zeta }^2+6 \lambda _1^{\Phi } v_{\Phi }^2+8 \lambda _3^{\Phi } v_{\Phi }^2)=0,\nonumber\\
\frac{\partial V}{\partial \Phi_3}&=& v_{\Phi }(3 v_H^2 \lambda _1^{\Phi H}+4 v_H^2 \lambda _3^{\Phi H}-\mu _{\Phi }^2+\lambda _1^{\Phi \Delta } v_{\Delta }^2-\nonumber\\&&\lambda _2^{\Phi \Delta } v_{\Delta }^2+\lambda _1^{\Phi \zeta } v_{\zeta }^2+6 \lambda _1^{\Phi } v_{\Phi }^2+8 \lambda _3^{\Phi } v_{\Phi }^2)=0,\nonumber\\
\frac{\partial V}{\partial \Delta_1}&=& 2 v_H^2 ((\lambda _3^{H\Delta}-\lambda _4^{\Delta }) v_{\Delta }+\mu_1)+2 (\lambda _3^{\Phi \Delta }-\lambda _4^{\Phi \Delta })\nonumber\\&& v_{\Delta } v_{\Phi }^2+3 v_1 \lambda _1^{\Phi \Delta } v_{\Phi }^2=0,\nonumber\\
\frac{\partial V}{\partial \Delta_2}&=& v_{\Phi }^2 (3 \lambda _1^{H\Delta} v_{\Delta }+2 \mu_1)+v_{\Delta } (-\mu _{\Delta }^2+3 v_H^2 \lambda _1^{\Phi \Delta }+\nonumber\\&&(2 \lambda _1^{\Delta }+\lambda _1^{\Delta \zeta } v_{\zeta }^2)=0,\nonumber\\
\frac{\partial V}{\partial \Delta_3}&=& 2 (v_H^2 ((\lambda _3^{H\Delta }+\lambda _4^{H\Delta }) v_{\Delta }+\mu_1)+(\lambda _3^{\Phi \Delta }+\lambda _4^{\Phi \Delta }) \nonumber\\&&v_{\Delta } v_{\Phi }^2)=0,\nonumber\\
\frac{\partial V}{\partial \zeta}&=& 2 v_{\zeta } (-\mu _{\zeta }^2+3 v_H^2 \lambda _1^{H\zeta}+\lambda _1^{\Delta \zeta } v_{\Delta }^2+2 \lambda ^{\zeta } v_{\zeta }^2+3 \lambda _1^{\Phi \zeta } \nonumber\\&&v_{\Phi }^2)=0.\nonumber
\end{eqnarray}

In a similar way, we can justify the chosen VEVs for the texture $M_{\nu}^2$.

\bibliographystyle{iopart-num}
\bibliography{ref.bib}

\end{document}